\documentclass[english]{llncs}
\usepackage[T1]{fontenc}
\usepackage{babel}

\makeatletter

\providecommand{\LyX}{L\kern-.1667em\lower.25em\hbox{Y}\kern-.125emX\@}
\newcommand{\lyxline}[1]{
  {#1 \vspace{1ex} \hrule width \columnwidth \vspace{1ex}}
}
\newcommand{\noun}[1]{\textsc{#1}}

\usepackage{latexsym}
\usepackage{amssymb}

\makeatother
\begin{document}

\vfill{}
\title{Combining Logic Programs and\\
Monadic Second Order Logics\\
 by Program Transformation}
\vfill{}

\author{Fabio Fioravanti{\small \( ^{1} \)}, Alberto Pettorossi{\small \( ^{2} \)},
Maurizio Proietti{\small \( ^{1} \) }}

\institute{(1) IASI-CNR, Viale Manzoni 30, I-00185 Roma, Italy \\
(2) DISP, University of Roma Tor Vergata, I-00133 Roma, Italy\texttt{}~\\
\texttt{\{fioravanti,adp,proietti\}@iasi.rm.cnr.it}}

\maketitle
{\centering \thispagestyle{plain}\par}

\begin{abstract}
We present a program synthesis method based on unfold/fold transformation
rules which can be used for deriving terminating definite logic programs
from formulas of the Weak Monadic Second Order theory of one successor
(WS1S). This synthesis method can also be used as a proof method which
is a decision procedure for closed formulas of WS1S. We apply our
synthesis method for translating CLP(WS1S) programs into logic programs
and we use it also as a proof method for verifying safety properties
of infinite state systems.
\end{abstract}

\section{Introduction\label{sec:intro}}

The Weak Monadic Second Order theories of \( k \) successors (WSkS)
are theories of the second order predicate logic which express properties
of finite sets of finite strings over a \( k \)-symbol alphabet (see
\cite{Tho97} for a survey). Their importance relies on the fact that
they are among the most expressive theories of predicate logic which
are decidable. These decidability results were proved in the 1960's
\cite{Buc60,ThW68}, but they were considered as purely theoretical
results, due to the very high complexity of the automata-based decision
procedures. 

In recent years, however, it has been shown that some Monadic Second
Order theories can, in fact, be decided by using ad-hoc, efficient
techniques, such as BDD's and algorithms for finite state automata.
In particular, the MONA system implements these techniques for the
WS1S and WS2S theories \cite{He&96}. 

The MONA system has been used for the verification of several non-trivial
finite state systems \cite{BaK98,Kl&96}. However, the Monadic Second
Order theories alone are not expressive enough to deal with properties
of infinite state systems and, thus, for the verification of such
systems alternative techniques have been used, such as those based
on the embedding of the Monadic Second Order theories into more powerful
logical frameworks (see, for instance, \cite{BaF00}).

In a previous paper of ours \cite{Fi&02a} we proposed a verification
method for infinite state systems based on CLP(WSkS), which is a constraint
logic programming language resulting from the embedding of WSkS into
logic programs. In order to perform proofs of properties of infinite
state systems in an automatic way according to the approach we have
proposed, we need a system for constraint logic programming which
uses a solver for WSkS formulas and, unfortunately, no such system
is available yet.

In order to overcome this difficulty, in this paper we propose a method
for translating CLP(WS1S) programs into logic programs. This translation
is performed by a two step program synthesis method which produces
terminating definite logic programs from WS1S formulas. Step 1 of
our synthesis method consists in deriving a normal logic program from
a WS1S formula, and it is based on a variant of the Lloyd-Topor transformation~\cite{Llo87}.
Step~2 consists in applying an unfold/fold transformation strategy
to the normal logic program derived at the end of Step~1, thereby
deriving a terminating definite logic program. Our synthesis method
follows the general approach presented in \cite{PeP00a,PeP02a}. We
leave it for future research the translation into logic programs starting
from general CLP(WSkS) programs.

The specific contributions of this paper are the following ones. 

(1) We provide a synthesis strategy which is guaranteed to terminate
for any given WS1S formula. 

(2) We prove that, when we start from a closed WS1S formula \( \varphi  \),
our synthesis strategy produces a program which is either (i) a unit
clause of the form \( f\leftarrow  \), where \( f \) is a nullary
predicate equivalent to the formula \( \varphi  \), or (ii) the empty
program. Since in case (i) \( \varphi  \) is true and in case (ii)
\( \varphi  \) is false, our strategy is also a decision procedure
for WS1S formulas. 

(3) We show through a non-trivial example, that our verification method
based on CLP(WS1S) programs is useful for verifying properties of
infinite state transition systems. In particular, we prove the safety
property of a mutual exclusion protocol for a set of processes whose
cardinality may change over time. Our verification method requires:
(i)~the encoding into WS1S formulas of both the transition relation
and the elementary properties of the states of a transition system,
and (ii)~the encoding into a CLP(WS1S) program of the safety property
under consideration. Here we perform our verification task by translating
the CLP(WS1S) program into a definite logic program, thereby avoiding
the use of a solver for WS1S formulas. The verification of the safety
property has been performed by using a prototype tool built on top
of the MAP transformation system~\cite{MAP}.

\section{The Weak Monadic Second Order Theory of One Successor\label{sec:WS1S}}

We will consider a \emph{first order} presentation of the Weak Monadic
Second Order theory of one successor (WS1S). This first order presentation
consists in writing formulas of the form \( n\in S \), where \( \in  \)
is a first order predicate symbol (to be interpreted as membership
of a natural number to a finite set of natural numbers), instead of
formulas of the form \( S(n) \), where \( S \) is a \emph{predicate}
\emph{variable} (to be interpreted as ranging over finite sets of
natural numbers). 

We use a \emph{typed} first order language, with the following two
types: \emph{nat}, denoting the set of natural numbers, and \emph{set},
denoting the set of the finite sets of natural numbers (for a brief
presentation of the typed first order logic the reader may look at
\cite{Llo87}). The alphabet of WS1S consists of: (i) a set \emph{\( \mathit{Ivars} \)}
of \emph{individual variables} \( N,N_{1},N_{2},\ldots  \) of type
\emph{nat}, (ii) a set \( \mathit{Svars} \) of \emph{set variables}
\( S,S_{1},S_{2},\ldots  \) of type \emph{set}, (iii) the nullary
function symbol \( 0 \) (\emph{zero}) of type \emph{nat}, and the
unary function symbol \( s \) (\emph{successor}) of type \( \mathit{nat}\rightarrow \mathit{nat} \),
and (iv) the binary predicate symbols \( \leq  \) of type \( \mathit{nat}\times \mathit{nat} \),
and \( \in  \) of type \( \mathit{nat}\times \mathit{set} \). \emph{\( \mathit{Ivars}\cup \mathit{Svars} \)}
is ranged over by \emph{}\( X,X_{1},X_{2},\ldots  \) The syntax of
WS1S is defined by the following grammar:

\smallskip{}
\begin{tabular}{ll}
\emph{Individual terms}:~~~&
\( n\, ::=\, 0\, \, \, |\, \, \, N\, \, \, |\, \, \, s(n) \)~\\
\emph{Atomic formulas}:&
\( A\, ::=\, n_{1}\! \leq \! n_{2}\, \, \, |\, \, \, n\! \in \! S \)\\
\emph{Formulas}:&
\( \varphi \, ::=\, A\, \, \, |\, \, \, \neg \varphi \, \, \, |\, \, \, \varphi _{1}\wedge \varphi _{2}\, \, \, |\, \, \, \exists N\, \varphi \, \, \, |\, \, \, \exists S\, \varphi  \)\\
\end{tabular}
\smallskip{}

\noindent When writing formulas we feel free to use also the connectives
\( \vee  \), \( \rightarrow  \), \( \leftrightarrow  \) and the
universal quantifier \( \forall  \), as shorthands of the corresponding
formulas with \( \neg  \), \( \wedge  \), and \( \exists  \). Given
any two individual terms \( n_{1} \) and \( n_{2} \), we will write
the formulas \( n_{1}\! =\! n_{2} \), \( n_{1}\! \neq \! n_{2} \),
and \( n_{1}\! <\! n_{2} \) as shorthands of the corresponding formulas
using \( \leq  \). Notice that, for reasons of simplicity, we have
assumed that the symbol \( \leq  \) is primitive, although it is
also possible to define it in terms of \( \in  \) \cite{Tho97}. 

An example of a WS1S formula is the following formula \( \mu  \),
with free variables \( N \) and \( S \), which expresses that \( N \)
is the maximum number in a finite set \( S \):
\smallskip{}

{\centering \( \mu \, :\, \, N\! \in \! S\wedge \neg \exists N_{1}(N_{1}\! \in \! S\wedge \neg N_{1}\! \leq \! N) \)\par}
\smallskip{}

\noindent The semantics of WS1S formulas is defined by considering
the following \emph{typed interpretation} \( \mathcal{N} \):
\smallskip{}

\noindent (i)~the domain of the type \( \mathit{nat} \) is the set
\( \mathit{Nat} \) of the natural numbers and the domain of the type
\( \mathit{set} \) is the set \( P_{\mathit{fin}}(\mathit{Nat}) \)
of all finite subsets of \emph{Nat}; 
\smallskip{}

\noindent (ii)~the constant symbol \( 0 \) is interpreted as the
natural number \( 0 \) and the function symbol \( s \) is interpreted
as the successor function from \( \mathit{Nat} \) to \( \mathit{Nat} \);
\smallskip{}

\noindent (iii)~the predicate symbol \( \leq  \) is interpreted
as the less-or-equal relation on natural numbers, and the predicate
symbol \( \in  \) is interpreted as the membership of a natural number
to a finite set of natural numbers. 
\smallskip{}

The notion of a \emph{variable assignment} \( \sigma  \) \emph{over
a typed interpretation} is analogous to the untyped case, except that
\( \sigma  \) assigns to a variable an element of the domain of the
type of the variable. The definition of the \emph{satisfaction relation}
\( I\models _{\sigma }\varphi  \), where \( I \) is a typed interpretation
and \( \sigma  \) is a variable assignment is also analogous to the
untyped case, with the only difference that when we interpret an existentially
quantified formula we assume that the quantified variable ranges over
the domain of its type. We say that a formula \( \varphi  \) is \emph{true}
in an interpretation \( I \), written as \( I\models \varphi  \),
iff \( I\models _{\sigma }\varphi  \) for all variable assignments
\( \sigma  \). The problem of checking whether or not a WS1S formula
is true in the interpretation \( \mathcal{N} \) is decidable \cite{Buc60}.

\section{Translating WS1S Formulas into Normal Logic Programs\label{sec:translating}}

In this section we illustrate Step 1 of our method for synthesizing
definite programs from WS1S formulas. In this step, starting from
a WS1S formula, we derive a \emph{stratified} normal logic program~\cite{ApB94}
(simply called \emph{stratified programs}) by applying a variant of
the Lloyd-Topor transformation, called \emph{typed Lloyd-Topor transformation}.
Given a stratified program \( P \), we denote by \( M(P) \) its
\emph{perfect model} (which is equal to its \emph{least Herbrand model}
if \( P \) is a definite program) \cite{ApB94}.

Before presenting the typed Lloyd-Topor transformation, we need to
introduce a definite program, called \emph{NatSet}, which axiomatizes:
(i)~the natural numbers, (ii)~the finite sets of natural numbers,
(iii)~the ordering on natural numbers (\( \leq  \)), and (iv)~the
membership of a natural number to a finite set of natural numbers
(\( \in  \)). We represent: (i)~a natural number \( k\, (\geq \! 0) \)
as a ground term of the form \( s^{k}(0) \), and (ii)~a set of natural
numbers as a finite, ground list \( [b_{0},b_{1},\ldots ,b_{m}] \)
where, for \( i=0,\ldots ,m \), we have that \( b_{i} \) is either
\( {\tt y} \) or \( {\tt n} \). A number \( k \) belongs to the
set represented by \( [b_{0},b_{1},\ldots ,b_{m}] \) iff \( b_{k}={\tt y} \).
Thus, the finite, ground lists \( [b_{0},b_{1},\ldots ,b_{m}] \)
and \( [b_{0},b_{1},\ldots ,b_{m},{\tt n},\ldots ,{\tt n}] \) represent
the same set. In particular, the empty set is represented by any list
of the form \( [{\tt n},\ldots ,{\tt n}] \). The program \emph{NatSet}
consists of the following clauses (we adopt infix notation for \( \leq  \)
and \( \in  \)):
\smallskip{}

{\noindent \centering \begin{tabular}{llccl}
~~&
\( \mathit{nat}(0)\leftarrow  \)&
~~~~~~~~~~~~~&
~~&
\( 0\! \leq \! N\leftarrow  \)\\
&
\( \mathit{nat}(s(N))\leftarrow \mathit{nat}(N) \)&
{\par}&
&
\( s(N_{1})\! \leq \! s(N_{2})\leftarrow N_{1}\! \leq \! N_{2} \)\\
&
\( \mathit{set}([\, ])\leftarrow  \) &
&
&
\( 0\! \in \! [{\tt y}|S]\leftarrow  \) \\
&
\( \mathit{set}([{\tt y}|S])\leftarrow \mathit{set}(S) \) &
&
&
\( s(N)\! \in \! [B|S]\leftarrow N\! \in \! S \) \\
&
\( \mathit{set}([{\tt n}|S])\leftarrow \mathit{set}(S) \) &
&
&
\\
\end{tabular}\par}
\smallskip{}

\noindent Atoms of the form \( \mathit{nat}(N) \) and \( \mathit{set}(S) \)
are called \emph{type} \emph{atoms}. Now we will establish a correspondence
between the set of WS1S formulas which are true in \( \mathcal{N} \)
and the set of the so-called \emph{explicitly typed} WS1S formulas
which are true in the least Herbrand model \( M(\mathit{Nat}Set) \)
(see Theorem~\ref{thm:WS1S_characterization} below). 

Given a WS1S formula \( \varphi  \), the \emph{explicitly typed}
WS1S formula corresponding to \( \varphi  \) is the formula \( \varphi _{\tau } \)
constructed as follows. We first replace the subformulas of the form
\( \exists N\, \psi  \) by \( \exists N\, (\mathit{nat}(N)\wedge \psi ) \)
and the subformulas of the form \( \exists S\, \psi  \) by \( \exists S\, (set(S)\wedge \psi ) \),
thereby getting a new formula \( \varphi _{\eta } \) where every
bound (individual or set) variable occurs in a type atom. Then, we
get:

{\centering \( \varphi _{\tau }\, :\, \, \mathit{nat}(N_{1})\wedge \ldots \wedge \mathit{nat}(N_{h})\wedge \mathit{set}(S_{1})\wedge \ldots \wedge \mathit{set}(S_{k})\wedge \varphi _{\eta } \) \par}
\smallskip{}

\noindent where \( N_{1},\ldots ,N_{h},S_{1},\ldots ,S_{k} \) are
the variables which occur free in \( \varphi  \). 

For instance, let us consider again the formula \( \mu  \) which
expresses that \( N \) is the maximum number in a set \( S \). The
explicitly typed formula corresponding to \( \mu  \) is the following
formula:
\smallskip{}

{\centering \( \mu _{\tau }\, :\, \, \mathit{nat}(N)\wedge \mathit{set}(S)\wedge N\! \in \! S\wedge \neg \exists N_{1}(\mathit{nat}(N_{1})\wedge N_{1}\! \in \! S\wedge \neg N_{1}\! \leq \! N) \)\par}
\smallskip{}

For reasons of simplicity, in the following Theorem~\ref{thm:WS1S_characterization}
we identify: (i) a natural number \( k\, (\geq 0) \) in \emph{Nat}
with the ground term \( s^{k}(0) \) representing that number, and
(ii) a finite set of natural numbers in \( P_{\mathit{fin}}(\mathit{Nat}) \)
with any finite, ground list representing that set. By using these
identifications, we can view any variable assignment over the typed
interpretation \( \mathcal{N} \) also as a variable assignment over
the untyped interpretation \( M(\mathit{Nat}Set) \) (but not vice
versa).

\begin{theorem}
\noindent \textup{\label{thm:WS1S_characterization} Let \( \varphi  \)
be a WS1S formula and let \( \varphi _{\tau } \) be the explicitly
typed formula corresponding to \( \varphi  \). For every variable
assignment \( \sigma  \) over \( \mathcal{N} \),\[
\mathcal{N}\models _{\sigma }\varphi \, \, \, \, \mbox {iff}\, \, \, \, M(\mathit{Nat}Set)\models _{\sigma }\varphi _{\tau }\]
}
\end{theorem}
\begin{proof}
The proof proceeds by induction on the structure of the formula \( \varphi  \). 

\noindent (i) Suppose that \( \varphi  \) is of the form \( n_{1}\! \leq \! n_{2} \).
By the definition of the satisfaction relation, \( \mathcal{N}\models _{\sigma }n_{1}\! \leq \! n_{2} \)
iff the natural number \( \sigma (n_{1}) \) is less or equal than
the natural number \( \sigma (n_{2}) \). By the definition of least
Herbrand model and by using the clauses in \emph{NatSet} which define
\( \leq  \), \( \sigma (n_{1}) \) is less or equal than \( \sigma (n_{2}) \)
iff \( M(\mathit{NatSet})\models \sigma (n_{1})\! \leq \! \sigma (n_{2}) \)
(here we identify every natural number \( n \) with the ground term
\( s^{n}(0) \)). It can be shown that \( M(\mathit{NatSet})\models \mathit{nat}(\sigma (n_{1})) \)
and \( M(\mathit{NatSet})\models \mathit{nat}(\sigma (n_{2})) \).
Thus, \( M(\mathit{NatSet})\models \sigma (n_{1})\! \leq \! \sigma (n_{2}) \)
iff \( M(\mathit{NatSet})\models _{\sigma }\mathit{nat}(n_{1})\wedge \mathit{nat}(n_{2})\wedge n_{1}\! \leq \! n_{2} \).
Now, the term \( n_{1} \) is either of the form \( s^{m1}(0) \)
or of the form \( s^{m1}(N_{1}) \), where \( m1 \) is a natural
number. Similarly, the term \( n_{2} \) is either of the form \( s^{m2}(0) \)
or of the form \( s^{m2}(N) \), where \( m2 \) is a natural number.
We consider the case where \( n_{1} \) is \( s^{m1}(N_{1}) \) and
\( n_{2} \) is \( s^{m2}(N_{2}) \). The other cases are similar
and we omit them. It can be shown that, for all natural numbers \( m \),
\( M(\mathit{NatSet})\models _{\sigma }\mathit{nat}(s^{m}(N)) \)
iff \( M(\mathit{NatSet})\models _{\sigma }\mathit{nat}(N) \). Thus,
\( M(\mathit{NatSet})\models _{\sigma }\mathit{nat}(s^{m1}(N_{1}))\wedge \mathit{nat}(s^{m2}(N_{2}))\wedge s^{m1}(N_{1})\! \leq \! s^{m2}(N_{2}) \)
iff \( M(\mathit{NatSet})\models _{\sigma }\mathit{nat}(N_{1})\wedge \mathit{nat}(N_{2})\wedge s^{m1}(N_{1})\! \leq \! s^{m2}(N_{2}) \),
that is, \( M(\mathit{NatSet})\models _{\sigma }(n_{1}\! \leq \! n_{2})_{\tau } \).

\noindent (ii) The case where \( \varphi  \) is of the form \( n\! \in S \)
is similar to Case (i). 

\noindent (iii) Suppose that \( \varphi  \) is of the form \( \neg \psi  \).
By the definition of the satisfaction relation and the induction hypothesis,
\( \mathcal{N}\models _{\sigma }\neg \psi  \) iff \( M(\mathit{NatSet})\models _{\sigma }\neg (\psi _{\tau }) \).
Since \( \psi _{\tau } \) is of the form \( a_{1}(X_{1})\wedge \ldots \wedge a_{k}(X_{k})\wedge \psi _{\eta } \),
where \( X_{1},\ldots ,X_{k} \) are the free variables in \( \psi  \)
and \( a_{1}(X_{1}),\ldots ,a_{k}(X_{k}) \) are type atoms, by logical
equivalence, we get: \( M(\mathit{NatSet})\models _{\sigma }\neg (\psi _{\tau }) \)
iff \( M(\mathit{NatSet})\models _{\sigma }(a_{1}(X_{1})\wedge \ldots \wedge a_{k}(X_{k})\wedge \neg (\psi _{\eta }))\vee \neg (a_{1}(X_{1})\wedge \ldots \wedge a_{k}(X_{k})) \).
Finally, since for all variable assignments \( \sigma  \), \( M(\mathit{NatSet})\models _{\sigma }a_{1}(X_{1})\wedge \ldots \wedge a_{k}(X_{k}) \),
we have that \( M(\mathit{NatSet})\models _{\sigma }\neg (\psi _{\tau }) \)
iff \( M(\mathit{NatSet})\models _{\sigma }(a_{1}(X_{1})\wedge \ldots \wedge a_{k}(X_{k})\wedge \neg (\psi _{\eta })) \),
that is, \( M(\mathit{NatSet})\models _{\sigma }(\neg \psi )_{\tau } \)
(to see this, note that \( \neg (\psi _{\eta }) \) is equal to \( (\neg \psi )_{\eta } \)). 

\noindent (iv) The case where \( \varphi  \) is of the form \( \psi _{1}\wedge \psi _{2} \)
is similar to Case (iii). 

\noindent (v) Suppose that \( \varphi  \) is of the form \( \exists N_{1}\, \psi  \).
By the definition of the satisfaction relation and by the induction
hypothesis, \( \mathcal{N}\models _{\sigma }\exists N_{1}\, \psi  \)
iff there exists \( n_{1} \) in \emph{Nat} such that \( M(\mathit{NatSet})\models _{\sigma [N_{1}\mapsto n_{1}]}\psi _{\tau } \).
Since \( \psi _{\tau } \) is of the form \( \mathit{nat}(N_{1})\wedge \ldots \wedge \mathit{nat}(N_{h})\wedge \mathit{set}(S_{1})\wedge \ldots \wedge \mathit{set}(S_{k})\wedge \psi _{\eta } \),
where \( N_{1},\ldots ,N_{h},S_{1},\ldots ,S_{k} \) are the free
variables in \( \psi  \), we have that:

\noindent there exists \( n_{1} \) in \emph{Nat} such that \( M(\mathit{NatSet})\models _{\sigma [N_{1}\mapsto n_{1}]}\psi _{\tau } \) 

\noindent iff \( M(\mathit{NatSet})\models _{\sigma }\exists N_{1}\, (\mathit{nat}(N_{1})\wedge \ldots \wedge \mathit{nat}(N_{h})\wedge \mathit{set}(S_{1})\wedge \ldots \wedge \mathit{set}(S_{k})\wedge \psi _{\eta }) \) 

\noindent iff (by logical equivalence) \( M(\mathit{NatSet})\models _{\sigma }\mathit{nat}(N_{2})\wedge \ldots \wedge \mathit{nat}(N_{h})\wedge \mathit{set}(S_{1})\wedge \ldots \wedge \mathit{set}(S_{k})\wedge (\exists N_{1}\, \mathit{nat}(N_{1})\wedge \psi _{\eta }) \)

\noindent iff (by definition of explicitly typed formula) \( M(\mathit{NatSet})\models _{\sigma }(\exists N_{1}\, \psi )_{\tau } \). 

\noindent (vi) The case where \( \varphi  \) is of the form \( \exists S\, \psi  \)
is similar to Case (v). \hfill\( \Box  \)
\end{proof}
As a straightforward consequence of Theorem~\ref{thm:WS1S_characterization},
we have the following result.

\begin{corollary}
\noindent \textup{For every closed WS1S formula \( \varphi  \), \( \mathcal{N}\models \varphi  \)
iff \( M(\mathit{Nat}Set)\models \varphi _{\tau } \).}
\end{corollary}
Notice that the introduction of type atoms is indeed necessary, because
there are WS1S formulas \( \varphi  \) such that \( \mathcal{N}\models \varphi  \)
and \( M(\mathit{Nat}Set)\not \models \varphi  \). For instance,
\( \mathcal{N}\models \forall N_{1}\exists N_{2}\, N_{1}\! \leq \! N_{2} \)
and \( M(\mathit{Nat}Set)\not \models \forall N_{1}\exists N_{2}\, N_{1}\! \leq \! N_{2} \).
Indeed, for a variable assignment \( \sigma  \) over \( M(\mathit{Nat}Set) \)
which assigns \( [\, ] \) to \( N_{1} \), we have \( M(\mathit{Nat}Set)\not \models _{\sigma }\exists N_{2}\, N_{1}\! \leq \! N_{2} \).
(Notice that \( \sigma  \) is not a variable assignment over \( \mathcal{N} \)
because \( [\, ] \) is not a natural number.)
\medskip{}

Now we present a variant of the method proposed by Lloyd and Topor~\cite{Llo87},
called \emph{typed Lloyd-Topor transformation}, which we use for deriving
a stratified program from a given WS1S formula \( \varphi  \). We
need to consider a class of formulas of the form: \( A\leftarrow \beta  \),
called \emph{statements}, where \( A \) is an atom, called the \emph{head}
of the statement, and \( \beta  \) is a formula of the first order
predicate calculus, called the \emph{body} of the statement. In what
follows we write \( C[\gamma ] \) to denote a formula where the subformula
\( \gamma  \) occurs as an \emph{outermost} \emph{conjunct}, that
is, \( C[\gamma ]\, \, =\, \, \psi _{1}\wedge \gamma \wedge \psi _{2} \)
for some subformulas \( \psi _{1} \) and \( \psi _{2} \). 
\medskip{}

\lyxline{\normalsize}\vspace{-1\parskip}
\noindent \textbf{The Typed Lloyd-Topor Transformation.}

\noindent We are given in input a set of statements, where: (i) we
assume without loss of generality, that the only connectives and quantifiers
occurring in the body of the statements are \( \neg ,\wedge , \)
and \( \exists  \), and (ii) \( X,X_{1},X_{2},\ldots  \) denote
either individual or set variables. 

\noindent We perform the following transformation (A) and then the
transformation (B): 

\medskip{}
\noindent (A) We repeatedly apply the following rules A.1--A.4 until
a set of clauses is generated:
\medskip{}

\noindent \begin{tabular}{lcl}
(A.1) \( A\leftarrow C[\neg \neg \gamma ] \)&
~~is replaced by~~&
\( A\leftarrow C[\gamma ] \).\\
\end{tabular}
\smallskip{}

\noindent \begin{tabular}{lcl}
(A.2) \( A\leftarrow C[\neg (\gamma \wedge \delta )] \)&
~~is replaced by~~&
\( A\leftarrow C[\neg \mathit{newp}(X_{1},\ldots ,\, X_{k})] \)\\
&
&
\( \mathit{newp}(X_{1},\ldots ,\, X_{k})\leftarrow \gamma \wedge \delta  \)\\
\end{tabular} 
\smallskip{}

\noindent where \emph{newp} is a new predicate and \( X_{1},\ldots ,\, X_{k} \)
are the variables which occur free in \( \gamma \wedge \delta  \). 
\smallskip{}

\noindent \begin{tabular}{lcl}
(A.3) \( A\leftarrow C[\neg \exists X\, \gamma ] \) &
~~is replaced by~~&
\( A\leftarrow C[\neg \mathit{newp}(X_{1},\ldots ,\, X_{k})] \)\\
&
&
\( \mathit{newp}(X_{1},\ldots ,\, X_{k})\leftarrow \gamma  \)\\
\end{tabular}
\smallskip{}

\noindent where \emph{newp} is a new predicate and \( X_{1},\ldots ,X_{k} \)
are the variables which occur free in \( \exists X\, \gamma  \).
\smallskip{}

\noindent \begin{tabular}{ccc}
(A.4) \( A\leftarrow C[\exists X\, \gamma ] \)&
~~is replaced by~~&
\( A\leftarrow C[\gamma \{X/X_{1}\}] \)\\
\end{tabular}
\smallskip{}

\noindent where \( X_{1} \) is a new variable.
\medskip{}

\noindent (B) Every clause \( A\leftarrow G \) is replaced by \( A\leftarrow G_{\tau } \).
\lyxline{\normalsize}

\medskip{}
\noindent Given a WS1S formula \( \varphi  \) with free variables
\( X_{1},\ldots ,X_{n} \), we denote by \( \mathit{Cls}(f,\varphi _{\tau }) \)
the set of clauses derived by applying the typed Lloyd-Topor transformation starting
from the singleton \( \{f(X_{1},\ldots ,X_{n})\leftarrow \varphi \} \),
where \( f \) is a new \( n \)-ary predicate symbol. By construction,
\( \mathit{Nat}Set\cup \mathit{Cls}(f,\varphi _{\tau }) \) is a stratified
program. We have the following theorem.

\begin{theorem}
\textup{\label{thm:corr_of_LT} Let \( \varphi  \) be a WS1S formula
with free variables \( X_{1},\ldots ,X_{n} \) and let \( \varphi _{\tau } \)
be the explicitly typed formula corresponding to \( \varphi  \).
For all ground terms \( t_{1},\ldots ,t_{n} \), we have that:}
\end{theorem}
{\noindent \centering \( M(\mathit{Nat}Set)\models \varphi _{\tau }\{X_{1}/t_{1},\ldots ,X_{n}/t_{n}\} \)~~iff \par}
\smallskip{}

{\noindent \centering \( M(\mathit{Nat}Set\cup \mathit{Cls}(f,\varphi _{\tau }))\models f(t_{1},\ldots ,t_{n}) \)\par}

\begin{proof}
It is similar to the proofs presented in \cite{Llo87,PeP00a} and
we omit it.
\end{proof}
From Theorems~\ref{thm:WS1S_characterization} and \ref{thm:corr_of_LT}
we have the following corollaries.
\smallskip{}

\begin{corollary}
\noindent \textup{For every WS1S formula \( \varphi  \) with free
variables \( X_{1},\ldots ,X_{n} \), and for every variable assignment
\( \sigma  \) over the typed interpretation \( \mathcal{N} \),\[
\mathcal{N}\models _{\sigma }\varphi \, \, \mbox {iff}\, \, M(\mathit{Nat}Set\cup \mathit{Cls}(f,\varphi _{\tau }))\models f(\sigma (X_{1}),\ldots ,\sigma (X_{n}))\]
}
\end{corollary}
\noindent %

\begin{corollary}
\noindent \textup{For every closed WS1S formula \( \varphi  \),  \[
\mathcal{N}\models \varphi \, \, \mbox {iff}\, \, M(\mathit{Nat}Set\cup \mathit{Cls}(f,\varphi _{\tau }))\models f\]
}
\end{corollary}
Let us consider again the formula \( \mu  \) we have considered above.
By applying the typed Lloyd-Topor transformation starting from the
singleton \( \{\mathit{max}(\mathit{S},N)\leftarrow \mu \} \) we
get the following set of clauses \( \mathit{Cls}(\mathit{max},\mu _{\tau }) \):
\smallskip{}

\( \mathit{max}(\mathit{S},N)\leftarrow \mathit{nat}(N)\wedge \mathit{set}(S)\wedge N\! \in \! S\wedge \neg \mathit{newp}(S,N) \)

\( \mathit{newp}(S,N)\leftarrow \mathit{nat}(N)\wedge \mathit{nat}(N_{1})\wedge \mathit{set}(S)\wedge N_{1}\! \in \! S\wedge \neg N_{1}\! \leq \! N \)

\smallskip{}
\noindent Unfortunately, the stratified program \( \mathit{Nat}Set\cup \mathit{Cls}(f,\varphi _{\tau }) \)
derived from the singleton \( \{f(X_{1},\ldots ,X_{n})\leftarrow \varphi \} \)
is not always satisfactory from a computational point of view because
it may not terminate when evaluating the query \( f(X_{1},\ldots ,X_{n}) \)
by using SLDNF resolution. (Actually, the above program \( \mathit{Cls}(\mathit{max},\mu _{\tau }) \)
which computes the maximum number of a set, terminates for all ground
queries, but in Section~\ref{sec:synthesis} we will give an example
where the program derived at the end of the typed Lloyd-Topor transformation
does not terminate.) Similar termination problems may occur by using
\emph{tabled resolution}~\cite{ChW96}, instead of SLDNF resolution.

To overcome this problem, we apply to the program \( \mathit{Nat}Set\cup \mathit{Cls}(f,\varphi _{\tau }) \)
the unfold/fold transformation strategy which we will describe in
Section~\ref{sec:synthesis}. In particular, by applying this strategy
we derive definite programs which terminate for all ground queries
by using LD resolution (that is, SLD resolution with the leftmost
selection rule).

\section{The Transformation Rules \label{sec:rules}}

In this section we describe the transformation rules which we use
for transforming stratified programs. These rules are a subset of
those presented in \cite{PeP00a,PeP02a}, and they are those required
for the unfold/fold transformation strategy presented in Section~\ref{sec:synthesis}. 

For presenting our rules we need the following notions. A variable
in the body of a clause \( C \) is said to be \emph{existential}
iff it does not occur in the head of \( C \). The \textit{definition}
of a predicate \( p \) in a program \( P \), denoted by \emph{\( \mathit{Def}(p,P) \)},
is the set of the clauses of \( P \) whose head predicate is \( p \).
The \textit{extended definition} of a predicate \( p \) in a program
\( P \), denoted by \emph{\( \mathit{Def}^{*}(p,P) \)}, is the union
of the definition of \( p \) and the definitions of all predicates
in \( P \) on which \( p \) depends. (See \cite{ApB94}for the definition
of the \emph{depends on} relation.) A program is \emph{propositional}
iff every predicate occurring in the program is nullary. Obviously,
if \( P \) is a propositional program then, for every predicate \( p \),
\( M(P)\models p \) is decidable.
\smallskip{}

A \emph{transformation sequence} is a sequence \( P_{0},\ldots ,P_{n} \)
of programs, where for \( 0\! \leq \! k\! \leq \! n\! -\! 1 \), program
\( P_{k+1} \) is derived from program \( P_{k} \) by the application
of one of the transformation rules R1--R4 listed below. For \( 0\! \leq \! k\! \leq \! n \),
we consider the set \( \mathit{Defs}_{k} \) of the clauses introduced
by the following rule R1 during the construction of the transformation
sequence \( P_{0},\ldots ,P_{k} \). 

When considering clauses of programs, we will feel free to apply the
following transformations which preserve the perfect model semantics:

(1) renaming of variables,

(2) rearrangement of the order of the literals in the body of a clause,
and

(3) replacement of a conjunction of literals the form \( L\wedge L \)
in the body of a clause by the literal \( L \).
\medskip{}

\medskip{}
\noindent \textbf{Rule R1. Definition.} We get the new program \( P_{k+1} \)
by adding to program \( P_{k} \) a clause of the form \( \mathit{newp}(X_{1},\ldots ,X_{r})\leftarrow L_{1}\wedge \ldots \wedge L_{m} \),
where: (i) the predicate \( \mathit{newp} \) is a predicate which
does not occur in \( P_{0}\cup \mathit{Defs}_{k} \), and (ii)~\( X_{1},\ldots ,X_{r} \)
are distinct (individual or set) variables occurring in \( L_{1}\wedge \ldots \wedge L_{m} \).
\medskip{}

\noindent \textbf{Rule R2. Unfolding.} Let \( C \) be a renamed apart
clause in \( P_{k} \) of the form: \( H\leftarrow G_{1}\wedge L\wedge G_{2} \),
where \( L \) is either the atom \( A \) or the negated atom \( \neg A \).
Let \( H_{1}\leftarrow B_{1},\ldots ,\, H_{m}\leftarrow B_{m} \),
with \( m\! \geq \! 0 \), be all clauses of program \( P_{k} \)
whose head is unifiable with \( A \) and, for \( j=1,\ldots ,m \),
let \( \vartheta _{j} \) the most general unifier of \( A \) and
\( H_{j} \). We consider the following two cases.
\smallskip{}

\noindent \emph{Case} 1: \( L \) is \( A \). By \textit{unfolding}
\textit{\emph{clause}} \textit{\( C \) w.r.t.~\( A \)} we derive
the new program \( P_{k+1}=(P_{k}-\{C\})\cup \{(H\leftarrow G_{1}\wedge B_{1}\wedge G_{2})\vartheta _{1},\ldots ,\, (H\leftarrow G_{1}\wedge B_{m}\wedge G_{2})\vartheta _{m}\} \).

\noindent In particular, if \( m\! =\! 0 \), that is, if we unfold
\( C \) w.r.t.~an atom which is not unifiable with the head of any
clause in \( P_{k} \), then we derive the program \( P_{k+1} \)
by deleting clause \( C \).
\smallskip{}

\noindent \emph{Case} 2: \( L \) is \( \neg A \). Assume that: (i)~\( A=H_{1}\vartheta _{1}=\cdots = \)
\( H_{m}\vartheta _{m} \), that is, for \( j=1,\ldots ,m \), \( A \)
is an instance of \( H_{j} \), (ii)~for \( j=1,\ldots ,m \), \( H_{j}\leftarrow B_{j} \)
has no existential variables, and (iii)~\( Q_{1}\vee \ldots \vee Q_{r} \),
with \( r\geq 0 \), is the disjunctive normal form of \( G_{1}\wedge \neg (B_{1}\vartheta _{1}\vee \ldots \vee B_{m}\vartheta _{m})\wedge G_{2} \).
By \textit{unfolding} \textit{\emph{clause}} \textit{\( C \) w.r.t.~\( \neg A \)}
we derive the new program \( P_{k+1}=(P_{k}-\{C\})\cup \{C_{1},\ldots ,\, C_{m}\} \),
where for \( j=1,\ldots ,r \), \( C_{j} \) is the clause \( H\leftarrow Q_{j} \).

\noindent In particular: (i) if \( m=0 \), that is, \( A \) is not
unifiable with the head of any clause in \( P_{k} \), then we get
the new program \( P_{k+1} \) by deleting \( \neg A \) from the
body of clause \( C \), and (ii) if for some \( j\in \{1,\ldots ,m\} \),
\( B_{j} \) is the empty conjunction, that is, \( A \) is an instance
of the head of a unit clause in \( P_{k} \), then we derive \( P_{k+1} \)
by deleting clause \( C \) from \( P_{k} \).
\medskip{}

\noindent \textbf{Rule R3. Folding.} Let \( C:\, H\leftarrow G_{1}\wedge B\vartheta \wedge G_{2} \)
be a renamed apart clause in \( P_{k} \) and \( D:\, \mathit{Newp}\leftarrow B \)
be a clause in \( \mathit{Defs}_{k} \). Suppose that for every existential
variable \( X \) of \( D \), we have that \( X\vartheta  \) is
a variable which occurs neither in \( \{H,G_{1},G_{2}\} \) nor in
the term \( Y\vartheta  \), for any variable \( Y \) occurring in
\( \mathit{B} \) and different from \( X \). By \textit{folding}
clause \( C \) using clause \( D \) we derive the new program \( P_{k+1}=(P_{k}-\{C\})\cup \{H\leftarrow G_{1}\wedge \mathit{Newp}\, \vartheta \wedge G_{2}\} \). 
\medskip{}

\noindent \textbf{Rule R4. Propositional Simplification.} Let \( p \)
be a predicate such that \( \mathit{Def}^{*}(p,P_{k}) \) is propositional.
If \( M(\mathit{Def}^{*}(p,P_{k}))\models p \) then we derive \( P_{k+1}=(P_{k}-\mathit{Def}(p,P_{k}))\cup \{p\leftarrow \} \).
If \( M(\mathit{Def}^{*}(p,P_{k}))\models \neg p \) then we derive
\( P_{k+1}=(P_{k}-\mathit{Def}(p,P_{k})) \).

\medskip{}
Notice that we can check whether or not \( M(P)\models p \) holds
by applying program transformation techniques \cite{PeP00a} and thus,
Rule R4 may be viewed as a derived rule.

The transformation rules R1--R4 we have introduced above, are collectively
called \emph{unfold/fold transformation rules}. We have the following
correctness result, similar to \cite{PeP00a}.

\begin{theorem}
\noindent \textbf{\emph{{[}Correctness of the Unfold/Fold Transformation
Rules{]}}} \label{thm:corr_of_rules} \textup{Let us assume that during
the construction of a transformation sequence \( P_{0},\ldots , \)
\( P_{n} \), each clause of \( \mathit{Defs}_{n} \) which is used
for folding, is unfolded (before or after its use for folding) w.r.t.~an
atom whose predicate symbol occurs in \( P_{0} \). Then, \[
M(P_{0}\cup \mathit{Defs}_{n})=M(P_{n}).\]
}
\end{theorem}
Notice that the statement obtained from Theorem \ref{thm:corr_of_rules}
by replacing `atom' by `literal', does not hold \cite{PeP00a}.

\section{The Unfold/Fold Synthesis Method\label{sec:synthesis}}

In this section we present our program synthesis method, called \emph{unfold/fold
synthesis} \emph{method}, which derives a definite program from any
given WS1S formula. We show that the synthesis method terminates for
all given formulas and also the derived programs terminate according
to the following notion of program termination: a program \( P \)
\emph{terminates for a query} \( Q \) iff every SLD-derivation of
\( P\cup \{\leftarrow Q\} \) via any computation rule is finite.

The following is an outline of our unfold/fold synthesis method.

\medskip{}\lyxline{\normalsize}\vspace{-1\parskip}
\noindent \textbf{The Unfold/Fold Synthesis} \textbf{\emph{}}\textbf{Method}\textbf{\emph{.}}

\noindent Let \( \varphi  \) be a WS1S formula with free variables
\( X_{1},\ldots ,X_{n} \) and let \( \varphi _{\tau } \) be the
explicitly typed formula corresponding to \( \varphi  \).
\smallskip{}

\noindent \emph{Step} 1\emph{.} We apply the \emph{typed Lloyd-Topor
transformation} and we derive a set \( \mathit{Cls}(f,\varphi _{\tau }) \)
of clauses such that: (i)~\( f \) is a new \( n \)-ary predicate
symbol, (ii)~\( \mathit{Nat}Set \) \( \cup \mathit{Cls}(f,\varphi _{\tau }) \)
is a stratified program, and (iii)~for all ground terms \( t_{1},\ldots ,t_{n} \),

\smallskip{}
\noindent (1)~\( M(\mathit{Nat}Set)\models \varphi _{\tau }\{X_{1}/t_{1},\ldots ,X_{n}/t_{n}\} \)~~iff

\noindent ~~~~~~~~~ \( M(\mathit{Nat}Set\cup \mathit{Cls}(f,\varphi _{\tau }))\models f(t_{1},\ldots ,t_{n}) \)
\smallskip{}

\noindent \emph{Step} 2\emph{.} We apply the \emph{unfold/fold transformation
strategy} (see below) and from the program \( \mathit{Nat}Set\cup \mathit{Cls}(f,\varphi _{\tau }) \)
we derive a definite program \( \mathit{TransfP} \) such that, for
all ground terms \( t_{1},\ldots ,t_{n} \),
\smallskip{}

\noindent (2.1)~~\( M(\mathit{Nat}Set\cup \mathit{Cls}(f,\varphi _{\tau }))\models f(t_{1},\ldots ,t_{n})\, \, \mbox {iff}\, \, M(\mathit{TransfP})\models f(t_{1},\ldots ,t_{n}) \);

\noindent (2.2)~~\( \mathit{TransfP} \) terminates for the query
\( f(t_{1},\ldots ,t_{n}) \).
\lyxline{\normalsize}\medskip{}

In order to present the unfold/fold transformation strategy which
we use for realizing Step~2 of our synthesis method, we introduce
the following notions of \emph{regular natset-typed clauses} and \emph{regular}
\emph{natset-typed definitions}. 

We say that a literal is \emph{linear} iff each variable occurs at
most once in it. The syntax of regular natset-typed clauses is defined
by the following grammar (recall that by \( N \) we denote individual
variables, by \( S \) we denote set variables, and by \( X,X_{1},X_{2},\ldots  \)
we denote either individual or set variables):
\smallskip{}

\begin{tabular}{lrll}
\emph{Head terms}:~~~&
\( h \)&
\( \, ::=\, \,  \)&
\( 0\, \, \, \, |\, \, \, \, s(N)\, \, \, \, |\, \, \, \, [\, ]\, \, \, \, |\, \, \, \, [{\tt y}|S]\, \, \, \, |\, \, \, \, [{\tt n}|S] \)\\
\emph{Clauses}:&
\( C \)&
\( \, ::=\, \,  \)&
\( p(h_{1},\ldots ,h_{k})\leftarrow \, \, \, \, |\, \, \, \, p_{1}(h_{1},\ldots ,h_{k})\leftarrow p_{2}(X_{1},\ldots ,X_{m}) \)\\
\end{tabular}
\smallskip{}

\noindent where for every clause \( C \), (i) both \( hd(C) \) and
\( bd(C) \) are linear atoms, and (ii)~\( \{X_{1},\ldots ,X_{m}\}\subseteq \mathit{vars}(h_{1},\ldots ,h_{k}) \)
(that is, \( C \) has no existential variables). A \emph{regular
natset-typed program} is a set of regular natset-typed clauses. 

The reader may check that the program \emph{NatSet} presented in Section~\ref{sec:translating}
is a regular natset-typed program. The following properties are straightforward
consequences of the definition of regular natset-typed program.

\begin{lemma}
\textup{\label{lemma:prop_of_natset}Let \( P \) be a regular natset-typed
program. Then:}

\noindent \textup{(i) \( P \) terminates for every ground query \( p(t_{1},\ldots ,t_{n}) \)
with \( n>0 \);}

\noindent \textup{(ii) If \( p \) is a nullary predicate then \( \mathit{Def}^{*}(p,P) \)
is propositional.}
\end{lemma}
The syntax of natset-typed definitions is given by the following grammar:

\smallskip{}
\begin{tabular}{lrll}
\emph{Individual terms}:~~~&
\( n \)&
\( \, ::=\, \,  \)&
\( 0\, \, \, \, |\, \, \, \, N\, \, \, \, |\, \, \, \, s(n) \)~\\
\emph{Terms}:&
\( t\,  \)&
\( \, ::=\, \,  \)&
\( n\, \, \, \, |\, \, \, \, S \)\\
\emph{Type atom}s:&
\( T \)&
\( \, ::=\, \,  \)&
\( \mathit{nat}(N)\, \, \, \, |\, \, \, \, \mathit{set}(S) \)\\
\emph{Literals}:&
\( L \)&
\( \, ::=\, \,  \)&
\( p(t_{1},\ldots ,t_{k})\, \, \, \, |\, \, \, \, \neg p(t_{1},\ldots ,t_{k}) \)\\
\emph{Definitions}:&
\raggedright \( D \)&
\( \, ::=\, \,  \)&
\( p(X_{1},\ldots ,X_{k})\leftarrow T_{1}\wedge \ldots \wedge T_{r}\wedge L_{1}\wedge \ldots \wedge L_{m} \)\\
\end{tabular}
\smallskip{}

\noindent where for all definitions \( D \), \( \mathit{vars}(D)\subseteq \mathit{vars}(T_{1}\wedge \ldots \wedge T_{r}) \). 

A sequence \( D_{1},\ldots ,D_{s} \) of natset-typed definitions
is said to be a \emph{hierarchy} iff for \( i=1,\ldots ,s \) the
predicate appearing in \( hd(D_{i}) \) does not occur in \( D_{1},\ldots ,D_{i-1},bd(D_{i}) \).
Notice that in a hierarchy of natset-typed definitions, any predicate
occurs in the head of at most one clause.

One can show that given a WS1S formula \( \varphi  \) the set \( \mathit{Cls}(f,\varphi _{\tau }) \)
of clauses derived by applying the typed Lloyd-Topor transformation
is a hierarchy \( D_{1},\ldots ,D_{s} \) of natset-typed definitions
and the last clause \( D_{s} \) is the one defining \( f \).
\newpage

\lyxline{\normalsize}\vspace{-1\parskip}
{\noindent \raggedright \textbf{The Unfold/Fold Transformation Strategy.} \par}

\smallskip{}
\noindent \emph{Input}: (i) A regular natset-typed program \( P \)
where for each nullary predicate \( p \), \( \mathit{Def}^{*}(p,\mathit{Transf}P) \)
is either the empty set or the singleton \( \{p\leftarrow \} \),
and (ii)~a hierarchy \( D_{1},\ldots ,D_{s} \) of natset-typed definitions
such that no predicate occurring in \( P \) occurs also in the head
of a clause in \( D_{1},\ldots ,D_{s} \).

\noindent \emph{Output}: A regular natset-typed program \( \mathit{TransfP} \)
such that, for all ground terms \( t_{1},\ldots ,t_{n} \),

\noindent (2.1)~~\( M(\mathit{P}\cup \{D_{1},\ldots ,D_{s}\})\models f(t_{1},\ldots ,t_{n}) \)
iff \( M(\mathit{TransfP})\models f(t_{1},\ldots ,t_{n}) \);

\noindent (2.2)~~\( \mathit{TransfP} \) terminates for the query
\( f(t_{1},\ldots ,t_{n}) \).

\noindent \underbar{~~~~~~~~~~~~~~~~~~~~~~~~~~~}

\smallskip{}
\noindent \( \mathit{TransfP}:=\mathit{P} \); \emph{\( \mathit{Defs}:=\emptyset  \)};
\smallskip{}

\noindent \textbf{\noun{for}} \( i=1,\ldots ,s \) \textbf{\noun{do}} \noun{}
\smallskip{}

\noindent \emph{\( \mathit{Defs}:=\mathit{Defs}\cup \{D_{i}\} \)};\noun{~~}~\( \mathit{InDefs}:=\{D_{i}\} \);

\noindent By the definition rule we derive the program \noun{\( \mathit{TransfP}\cup \mathit{InDefs} \).}
\medskip{}

\noindent \textbf{\noun{while}} \emph{\( \mathit{InDefs}\neq \emptyset  \)}
\textbf{\noun{do}}
\smallskip{}

\noindent (1)\emph{~Unfolding.} From \noun{}program \noun{}\( \mathit{TransfP}\cup \mathit{InDefs} \)
we derive \( \mathit{TransfP}\cup \mathit{U} \) \noun{}by: (i)~applying
the unfolding rule w.r.t.~each atom occurring positively in the body
of a clause in \( \mathit{InDefs} \), thereby deriving \noun{\( \mathit{TransfP}\cup \mathit{U}_{1} \)},
\noun{}then (ii)~applying the unfolding rule w.r.t.~each negative
literal occurring in the body of a clause in \( \mathit{U}_{1} \),
thereby deriving \noun{\( \mathit{TransfP}\cup \mathit{U}_{2} \)},
and, finally, (iii)~applying the unfolding rule w.r.t.~ground literals
until we derive a program \( \mathit{TransfP}\cup \mathit{U} \) such
that no ground literal occurs in the body of a clause of \( U \). 
\smallskip{}

\noindent (2)\emph{~Definition-Folding.} From program \noun{\( \mathit{TransfP}\cup U \)}
we derive \noun{}\( \mathit{TransfP}\cup \mathit{F}\cup \mathit{NewDefs} \)
as follows. Initially, \emph{NewDefs} is the empty set. For each non-unit
clause \( C \): \( H\leftarrow B \) in \( U \),\\
(i) we apply the definition rule and we add to \emph{NewDefs} a clause
of the form \( \mathit{newp}(X_{1},\ldots ,X_{k})\leftarrow B \),
where \( X_{1},\ldots ,X_{k} \) are the non-existential variables
occurring in \( B \), unless a variant clause already occurs in \emph{Defs},
modulo the head predicate symbol and the order and multiplicity of
the literals in the body, and\\
(ii) we replace \( C \) by the clause derived by folding \( C \)
w.r.t.~\( B \). The folded clause is an element of \( F \).\\
No transformation rule is applied to the unit clauses occurring in
\( U \) and, therefore, also these clauses are elements of \( F \).
\smallskip{}

\noindent (3)~\( \mathit{TransfP}:=\mathit{TransfP}\cup F \);\noun{~~}\emph{~\( \mathit{Defs}:=\mathit{Defs}\cup \mathit{NewDefs} \)};\noun{~~}~\( \mathit{InDefs}:=\mathit{NewDefs} \) 
\smallskip{}

\noindent \textbf{\emph{\noun{end}}} \textbf{\noun{while}}\noun{;}
\smallskip{}

\noindent \emph{Propositional Simplification.} For each predicate
\( p \) such that \( \mathit{Def}^{*}(p,\mathit{TransfP}) \) is
propositional, we apply the propositional simplification rule and 

\noindent \emph{if} \( M(\mathit{TransfP})\models p \) 

\noindent \emph{then} \( \mathit{TransfP}:=(\mathit{TransfP}-\mathit{Def}(p,\mathit{TransfP}))\cup \{p\leftarrow \} \)

\noindent \emph{else} \( \mathit{TransfP}:=(\mathit{TransfP}-\mathit{Def}(p,\mathit{TransfP})) \)
\medskip{}

\noindent \textbf{\emph{\noun{end for}}}
\lyxline{\normalsize}\medskip{}

The reader may verify that if we apply the unfold/fold transformation
strategy starting from the program \( \mathit{Nat}Set \) together
with the clauses \( \mathit{Cls}(\mathit{max},\mu _{\tau }) \) which
we have derived above by applying the typed Lloyd-Topor transformation,
we get the following final program:

\smallskip{}
\( \mathit{max}([{\tt y}|S],0)\leftarrow \mathit{new}1(S) \)

\( \mathit{max}([{\tt y}|S],s(N))\leftarrow \mathit{max}(S,N) \)

\( \mathit{max}([{\tt n}|S],s(N))\leftarrow \mathit{max}(S,N) \)

\( \mathit{new}1([\, ])\leftarrow  \)

\( \mathit{new}1([{\tt n}|S])\leftarrow \mathit{new}1(S) \)
\smallskip{}

\noindent To understand the first clause, recall that the empty set
is represented by any list of the form \( [{\tt n},\ldots ,{\tt n}] \).
A more detailed example of application of the unfold/fold transformation
strategy will be given later.

In order to prove the correctness and the termination of our unfold/fold
transformation strategy we need the following lemmas whose proofs
are mutually dependent.

\begin{lemma}
\textup{\label{lemma:regular} During the application of the unfold/fold
transformation strategy, \( \mathit{TransfP} \) is a regular natset-typed
program.}
\end{lemma}
\begin{proof}
Initially, \( \mathit{TransfP} \) is the regular natset-typed program
\( P \). Now we assume that \( \mathit{TransfP} \) is a regular
natset-typed program and we show that after an execution of the body
of the \noun{for} statement, \( \mathit{TransfP} \) is a regular
natset-typed program.

First we prove that after the execution of the \noun{while} statement,
\( \mathit{TransfP} \) is a regular natset-typed program. In order
to prove this, we show that every new clause \( E \) which is added
to \( \mathit{TransfP} \) at Point~(3) of the strategy is a regular
natset-typed clause. 

Clause \( E \) is derived from a clause \( D \) of \( \mathit{InDefs} \)
by unfolding (according to the Unfolding phase) and by folding (according
to the Definition-Folding phase). By Lemma~\ref{lemma:definitions},
\( D \) is a natset-typed definition of the form \( \mathit{p}(X_{1},\ldots ,X_{k})\leftarrow T_{1}\wedge \ldots \wedge T_{r}\wedge L_{1}\wedge \ldots \wedge L_{m} \).
By unfolding w.r.t.~the type atoms \( T_{1},\ldots ,T_{r} \) (according
to Point~(i) of the Unfolding phase) we get clauses of the form \( \mathit{p}(h_{1},\ldots ,h_{k})\leftarrow T'_{1}\wedge \ldots \wedge T'_{r1}\wedge L'_{1}\wedge \ldots \wedge L'_{m} \),
where: (a)~\( h_{1},\ldots ,h_{k} \) are head terms, (b)~\( \mathit{p}(h_{1},\ldots ,h_{k}) \)
is a linear atom (because \( X_{1},\ldots ,X_{k} \) are distinct
variables), and (c)~for \( i=1,\ldots ,m \), no argument of \( L'_{i} \)
is a variable. By the inductive hypothesis \( \mathit{TransfP} \)
is a regular natset-typed program and, therefore, by unfolding w.r.t.~the
literals \( L'_{1},\ldots ,L'_{m} \) (according to Points~(ii) and~(iii)
of the Unfolding phase) we get clauses of the form \( D': \) \( \mathit{p}(h_{1},\ldots ,h_{k})\leftarrow T'_{1}\wedge \ldots \wedge T'_{r1}\wedge L''_{1}\wedge \ldots \wedge L''_{m1} \).
Either \( D' \) is a unit clause or, by folding according to the
Definition-Folding phase, it is replaced by \( \mathit{p}(h_{1},\ldots ,h_{k})\leftarrow \mathit{newp}(X_{1},\ldots ,X_{m}) \)
where \( X_{1},\ldots ,X_{m} \) are the distinct, non-existential
variables occurring in \( bd(D') \). Hence, \( E \) is either a
unit clause of the form \( \mathit{p}(h_{1},\ldots ,h_{k})\leftarrow  \)
or a clause of the form \( \mathit{p}(h_{1},\ldots ,h_{k})\leftarrow \mathit{newp}(X_{1},\ldots ,X_{m}) \),
where \( \{X_{1},\ldots ,X_{m}\}\subseteq \mathit{vars}(h_{1},\ldots ,h_{k}) \).
Thus, \( E \) is a regular natset-typed clause. 

We conclude the proof by observing that if we apply the propositional
simplification rule to a natset-typed program, then we derive a natset-typed
program, because by this rule we can only delete clauses or add natset-typed
clauses of the form \( p\leftarrow  \). Thus, after an execution
of the body of the \noun{for} statement, \( \mathit{TransfP} \)
is a regular natset-typed program. \hfill$\Box$
\end{proof}
\begin{lemma}
\textup{\label{lemma:definitions} During the application of the unfold/fold
transformation strategy, \( \mathit{InDefs} \) is a set} \textup{\emph{}}\textup{of}
\textup{\emph{}}\textup{natset-typed definitions.}
\end{lemma}
\begin{proof}
Let us consider the \( i \)-th execution of the body of the \noun{for}
statement. Initially, \( \mathit{InDefs} \) is the singleton set
\( \{D_{i}\} \) of \emph{}natset-typed definitions. Now we assume
that \( \mathit{InDefs} \) is a set \emph{}of \emph{}natset-typed
definitions and we prove that, after an execution of the \noun{while}
statement, \( \mathit{InDefs} \) is a set of natset-typed definitions.
It is enough to show that every new clause \( E \) which is added
to \( \mathit{InDefs} \) at Point~(3) of the strategy, is a natset-typed
definition. By the Folding phase of the strategy, \( E \) is a clause
of the form \( \mathit{newp}(X_{1},\ldots ,X_{k})\leftarrow B \)
where \( B \) is the body of a clause derived from a clause \( D \)
of \( \mathit{InDefs} \) by unfolding. By the inductive hypothesis,
\( D \) is a natset-typed definition of the form \( \mathit{p}(X_{1},\ldots ,X_{k})\leftarrow T_{1}\wedge \ldots \wedge T_{r}\wedge L_{1}\wedge \ldots \wedge L_{m} \).
By unfolding w.r.t.~the type atoms \( T_{1},\ldots ,T_{r} \) (according
to Point~(i) of the Unfolding phase) we get clauses of the form \( D': \)
\( \mathit{p}(h_{1},\ldots ,h_{k})\leftarrow T'_{1}\wedge \ldots \wedge T'_{r1}\wedge L'_{1}\wedge \ldots \wedge L'_{m} \),
where \( \mathit{vars}(D')\subseteq \mathit{vars}(T'_{1}\wedge \ldots \wedge T'_{r1}) \).
Since, by Lemma~\ref{lemma:regular}, \( \mathit{TransfP} \) is
a regular natset-typed program, by unfolding w.r.t.~the literals
\( L'_{1},\ldots ,L'_{m} \) (according to Points~(ii) and~(iii)
of the Unfolding phase) we get clauses of the form \( D'': \) \( \mathit{p}(h_{1},\ldots ,h_{k})\leftarrow T'_{1}\wedge \ldots \wedge T'_{r1}\wedge L''_{1}\wedge \ldots \wedge L''_{m1} \)
where \( \mathit{vars}(D'')\subseteq \mathit{vars}(T'_{1}\wedge \ldots \wedge T'_{r1}) \).
Thus, \( E \) is a natset-typed definition of the form \( \mathit{newp}(X_{1},\ldots ,X_{k})\leftarrow T'_{1}\wedge \ldots \wedge T'_{r1}\wedge L''_{1}\wedge \ldots \wedge L''_{m1} \)
with \( \mathit{vars}(E)\subseteq \mathit{vars}(T'_{1}\wedge \ldots \wedge T'_{r1}) \). 

We conclude the proof by observing that the Propositional Simplification
phase does not change \( \mathit{InDefs} \), and thus, after the
execution of the body of the \noun{for} statement, \( \mathit{InDefs} \)
is a set of \emph{}natset-typed definitions. \hfill$\Box$
\end{proof}
\begin{theorem}
\textup{\label{thm:corr_of_strategy}Let \( P \) and \( D_{1},\ldots ,D_{s} \)
be the input program and the input hierarchy, respectively, of the
unfold/fold transformation strategy and let \( \mathit{TransfP} \)
be the output of the strategy. Then,}
\smallskip{}

\noindent \textup{(1)} \textup{\emph{TransfP}} \textup{is a natset-typed
program; }
\smallskip{}

\noindent \textup{(2) for every nullary predicate \( p \), \( \mathit{Def}^{*}(p,\mathit{TransfP}) \)
is either \( \emptyset  \) or \( \{p\leftarrow \} \); }
\smallskip{}

\noindent \textup{(3) for all ground terms \( t_{1},\ldots ,t_{n} \),}
\smallskip{}

\noindent \textup{~~(3.1)~~\( M(\mathit{P}\cup \{D_{1},\ldots ,D_{s}\})\models f(t_{1},\ldots ,t_{n}) \)
iff \( M(\mathit{TransfP})\models f(t_{1},\ldots ,t_{n}) \);}

\noindent \textup{~~(3.2)~~\( \mathit{TransfP} \) terminates
for the query \( f(t_{1},\ldots ,t_{n}) \).}
\end{theorem}
\begin{proof}
Point (1) is a straightforward consequence of Lemma~\ref{lemma:regular}. 
\smallskip{}

For Point (2), let us notice that, by Lemma~\ref{lemma:regular},
at each point of the unfold/fold transformation strategy \( \mathit{TransfP} \)
is a natset-typed program and therefore, by Lemma~\ref{lemma:prop_of_natset},
for every nullary predicate \( p \), \( \mathit{Def}^{*}(p,\mathit{TransfP}) \)
is propositional. Since the last step of the unfold/fold transformation
strategy consists in applying to \( \mathit{TransfP} \) the propositional
simplification rule for each predicate having a propositional extended
definition, \( \mathit{Def}^{*}(p,\mathit{TransfP}) \) is either
\( \emptyset  \) or \( \{p\leftarrow \} \).
\smallskip{}

Point (3.1) will be proved by using the correctness of the transformation
rules w.r.t. the Perfect Model semantics (see Theorem~\ref{thm:corr_of_rules}).
Let us first notice that the unfold/fold transformation strategy generates
a transformation sequence (see Section~\ref{sec:rules}), where:
the initial program is \( P \), the final program is the final value
of \( \mathit{TransfP} \), and the set of clauses introduced by the
definition rule R1 is the final value of \( \mathit{Defs} \).

To see that our strategy indeed generates a transformation sequence,
let us observe the following facts (A) and (B):
\smallskip{}

\noindent (A) The addition of \( \mathit{InDefs} \) to \( \mathit{TransfP} \)
at the beginning of each execution of the body of the \noun{for}
statement is an application of the \noun{}definition rule. Indeed,
for \( i=1,\ldots s, \) \( \mathit{InDefs}=\{D_{i}\} \) and, by
the hypotheses on the input sequence \( D_{1},\ldots ,D_{s} \), we
have that the head predicate of \( D_{i} \) does not occur in the
current value of \( P\cup \mathit{Defs} \).
\smallskip{}

\noindent (B) When we unfold the clauses of \( U_{1} \) w.r.t.~negative
literals, we have that:

\noindent (B.1) Condition (i) of Case (2) of the unfolding rule (see
Section~\ref{sec:rules}) is satisfied because:

\noindent (a)~Every clause \( D \) of \( \mathit{InDefs} \) is
a natset-typed definition (see Lemma~\ref{lemma:definitions}) and,
thus, for each variable \( X \) occurring in \( D \) there is a
type atom of the form \( a(X) \) in \( bd(D) \). Since we unfold
the clauses of \( \mathit{InDefs} \) w.r.t.~all the atoms which
occur positively in the bodies of the clauses in \( \mathit{InDefs} \),
and in particular, w.r.t.~type atoms, every argument of a negative
literal in the body of a clause of \( U_{1} \) is of one of the following
forms: \( 0 \), \( s(n) \), \( [\, ] \), \( [{\tt y}|S] \), \( [{\tt n}|S] \).

\noindent (b) For each negative literal \( \neg p(t_{1},\ldots ,t_{k}) \)
in the body of a clause of \( U_{1} \), the \textit{\emph{definition}}
of \( p \) is a subset of the regular natset-typed program \( \mathit{TransfP} \)
(see Lemma~\ref{lemma:regular}) and, hence, the head of a clause
in \( \mathit{TransfP} \) is a linear atom of the form \( p(h_{1},\ldots ,h_{k}) \),
where \( h_{1},\ldots ,h_{k} \) are head terms (see the definition
of regular natset-typed clauses above). 

\noindent From (a) and (b) it follows that if \( p(t_{1},\ldots ,t_{k}) \)
is unifiable with \( p(h_{1},\ldots ,h_{k}) \) then \( p(t_{1},\ldots ,t_{k}) \)
is an instance of \( p(h_{1},\ldots ,h_{k}) \).

\noindent (B.2) Condition (ii) of Case (2) of the unfolding rule is
satisfied because \( \mathit{TransfP} \) is a regular natset-typed
program (see Lemma~\ref{lemma:regular}) and, thus, no clause in
\( \mathit{TransfP} \) has existential variables. 
\smallskip{}

Now, the transformation sequence constructed by the unfold/fold transformation
strategy satisfies the hypothesis of Theorem~\ref{thm:corr_of_rules}.
Indeed, let us consider a clause \( D \) which is used for folding
a clause \( C \). Since \( C \) has been derived at the end of the
Unfolding phase, no ground literal occurs in \( bd(C) \) and, thus,
there is at least one variable occurring in \( D \). Hence, there
is at least one type atom in \( bd(D) \), because \( D \) is a natset-typed
definition (see Lemma~\ref{lemma:definitions}). Therefore, during
an application of the unfold/fold transformation strategy (before
or after the use of \( D \) for folding), \( D \) is unfolded w.r.t.~a
type atom (see Point~(i) of the Unfolding phase). Thus, by Theorem~\ref{thm:corr_of_rules},
we have that \( M(\mathit{P}\cup \mathit{Defs})=M(\mathit{TransfP}) \),
where by \( \mathit{Defs} \) and \( \mathit{TransfP} \) we indicate
the values of these variables at the end of the unfold/fold transformation
strategy. Observe that \( \mathit{Def}^{*}(f,\mathit{P}\cup \mathit{Defs})=\mathit{Def}^{*}(f,\mathit{P}\cup \{D_{1},\ldots ,D_{s}\}) \)
and, therefore, \( M(\mathit{P}\cup \{D_{1},\ldots ,D_{s}\})\models f(t_{1},\ldots ,t_{n}) \)
iff \( M(\mathit{P}\cup \mathit{Defs})\models f(t_{1},\ldots ,t_{n}) \)
iff \( M(\mathit{TransfP})\models f(t_{1},\ldots ,t_{n}) \). 

Finally, let us prove Point (3.2). We consider the following two cases:
\smallskip{}

\noindent (\( n=0 \)) \( f \) is nullary and hence, by Point (2)
of this theorem, \( \mathit{Def}^{*}(f,\mathit{TransfP}) \) is either
\( \emptyset  \) or \( \{f\leftarrow \} \). Thus, \emph{TransfP}
terminates for the query \( f \).
\smallskip{}

\noindent (\( n>0 \)) By Point (1) of this theorem, \emph{TransfP}
is a natset-typed program and thus, by Lemma~\ref{lemma:prop_of_natset},
\( \mathit{TransfP} \) terminates for the ground query \( f(t_{1},\ldots ,t_{n}) \). \hfill$\Box$
\end{proof}
\begin{theorem}
\textup{The unfold/fold transformation strategy terminates.}
\end{theorem}
\begin{proof}
We have to show that the \noun{while} statement in the body of the
\noun{for} statement terminates. 

Each execution of the Unfolding phase terminates. Indeed, (a)~the
number of applications of the unfolding rule at Points (i) and (ii)
is finite, because \emph{InDefs} is a finite set of clauses and the
body of each clause has a finite number of literals, and (b)~at Point
(iii) only a finite number of unfolding steps can be applied w.r.t.~ground
literals, because the program held by \( \mathit{TransfP} \) during
the Unfolding phase terminates for every ground query. To see this
latter fact, let us notice that, by Lemma~\ref{lemma:regular}, \( \mathit{TransfP} \)
is a natset-typed program. Thus, by Lemma~\ref{lemma:prop_of_natset},
\( \mathit{TransfP} \) terminates for any ground query \( p(t_{1},\ldots ,t_{n}) \)
with \( n\geq 1 \). For a ground query \( p \), where \( p \) is
a nullary predicate, \( \mathit{TransfP} \) terminates because \( \mathit{Def}^{*}(p,\mathit{Transf}P) \)
is either the empty set or it is the singleton \( \{p\leftarrow \} \).
Indeed, this follows from our assumptions on the input program and
from the execution of the Propositional Simplification phase after
completion of the \noun{while} statement. 

Each execution of the Definition-Folding phase terminates because
a finite number of clauses are introduced by definition and a finite
number of clauses are folded.

Thus, in order to show that the strategy terminates, it is enough
to show that after a finite number of executions of the body of the
\noun{while} statement, we get \( \mathit{InDefs}=\emptyset  \).
Let \( \mathit{Defs}_{j} \) and \( \mathit{InDefs}_{j} \) be the
values of \( \mathit{Defs} \) and \( \mathit{InDefs} \), respectively,
at the end of the \( j \)-th execution of the body of the \noun{while}
statement. If the \noun{while} statement terminates after \( z \)
executions of its body, then, for all \( j>z \), we define \( \mathit{Defs}_{j} \)
to be \( \mathit{Defs}_{z} \) and \( \mathit{InDefs}_{j} \) to be
\( \emptyset  \). We have that, for any \( j\geq 1 \), \( \mathit{InDefs}_{j}=\emptyset  \)
iff \( \mathit{Defs}_{j-1}=\mathit{Defs}_{j} \). Since for all \( j\geq 1 \),
\( \mathit{Defs}_{j-1}\subseteq \mathit{Defs}_{j} \), the termination
of the strategy will follow from the following property:
\smallskip{}

~~~~~~there exists \( K>0 \) such that, for all \( j\geq 1 \),
\( |\mathit{Defs}_{j}|\leq K \)~~~~~~({*})
\smallskip{}

\noindent Let \( \mathit{TransfP}_{0} \), \( \mathit{Defs}_{0} \),
and \( \mathit{InDefs}_{0}\, (\subseteq \mathit{Defs}_{0}) \) be
the values of \( \mathit{TransfP} \), \( \mathit{Defs} \), and \( \mathit{InDefs} \),
respectively, at the beginning of the execution of the \noun{while}
statement. By Lemma~\ref{lemma:definitions}, for all \( j\geq 1 \),
\( \mathit{Defs}_{j} \) is a set of natset-typed definitions. Property~({*})
follows from the fact that, for all \( D\in \mathit{Defs}_{j} \),
the following holds:

\noindent (a) every predicate occurring in \( bd(D) \) also occurs
in \( \mathit{TransfP}_{0}\cup \mathit{InDefs}_{0} \);

\noindent (b) for every literal \( L \) occurring in \( bd(D) \), 

\noindent ~~~~~~\( \mathit{height}(L)\leq \max \{\mathit{height}(M)\, |\, M \)
is a literal in the body of a clause in \( \mathit{Defs}_{0}\} \)

\noindent where the \emph{height} of a literal is defined as the length
of the maximal path from the root to a leaf of the literal considered
as a tree;

\noindent (c) \( |\mathit{vars}(D)|\leq \max \{\mathit{vars}(D')\, |\, D' \)
is a clause in \( \mathit{Defs}_{0}\} \); 

\noindent (d) no two clauses in \( \mathit{Defs}_{j} \) can be made
equal by one or more applications of the following transformations:
renaming of variables, renaming of head predicates, rearrangement
of the order of the literals in the body, and deletion of duplicate
literals.

\noindent Recall that \( bd(D) \) is equal to \( bd(E') \) where
\( E' \) is derived by unfolding (according to the Unfolding phase
of the strategy) a clause \( E \) in \( \mathit{TransfP}_{0}\cup \mathit{InDefs}_{j} \)
and \( E \) belongs to \( \mathit{InDefs}_{j} \).

\noindent Now Property (a) is a straightforward consequence of the
definition of the unfolding rule.

\noindent Property (b) can be shown as follows. \( E \) is of the
form \( \mathit{newp}(X_{1},\ldots ,X_{k})\leftarrow T_{1}\wedge \ldots \wedge T_{r}\wedge L_{1}\wedge \ldots \wedge L_{m} \).
By unfolding w.r.t.~the type atoms \( T_{1},\ldots ,T_{r} \) (according
to Point~(i) of the Unfolding phase) we get clauses of the form \( \mathit{newp}(h_{1},\ldots ,h_{k})\leftarrow T'_{1}\wedge \ldots \wedge T'_{r1}\wedge L'_{1}\wedge \ldots \wedge L'_{m} \),
where \( h_{1},\ldots ,h_{k} \) are head terms and, for all \( i\in \{1,\ldots ,m\} \),
\( \mathit{height}(L'_{i})\leq \mathit{height}(L_{i})+1 \). By Lemma~\ref{lemma:regular},
\( \mathit{TransfP}_{0} \) is a regular natset-typed program and,
therefore, by unfolding w.r.t.~the literals \( L'_{1},\ldots ,L'_{m} \)
(according to Point~(ii) of the Unfolding phase) we get clauses of
the form \( \mathit{newp}(h_{1},\ldots ,h_{k})\leftarrow T'_{1}\wedge \ldots \wedge T'_{r1}\wedge L''_{1}\wedge \ldots \wedge L''_{m1} \),
where for all \( i\in \{1,\ldots ,m1\} \), there exists \( i1\in \{1,\ldots ,m\}, \)
such that \( \mathit{height}(L''_{i})=\mathit{height}(L'_{i1})-1 \).
Thus, Property (b) follows from the fact that \( E' \) is derived
by unfolding w.r.t.~ground literals from a clause of the form \( \mathit{newp}(h_{1},\ldots ,h_{k})\leftarrow T'_{1}\wedge \ldots \wedge T'_{r1}\wedge L''_{1}\wedge \ldots \wedge L''_{m1} \)
and every unfolding w.r.t.~a ground literal does not increase the
height of the other literals in a clause.

\noindent Property (c) follows from Lemma~\ref{lemma:regular} and
the fact that by unfolding a clause \( E \) using regular natset-typed
clauses we get clauses \( E' \) where \( \mathit{vars}(E')\subseteq \mathit{vars}(E) \).
To see this, recall that in a regular natset-typed clause \( C \)
every term has at most one variable and \( \mathit{vars}(bd(C))\subseteq \mathit{vars}(hd(C)) \)
and, thus, by unfolding, a variable is replaced by a term with at
most one variable and no new variables are introduced. 

\noindent Finally, Point (d) is a consequence of Point (i) of the
Definition-Folding phase of the unfold/fold strategy. \hfill$\Box$
\end{proof}

\section{Deciding WS1S via the Unfold/Fold Proof Method}

In this section we show that if we start from a \emph{closed} WS1S
formula \( \varphi  \), our synthesis method can be used for checking
whether or not \( \mathcal{N}\models \varphi  \) holds and, thus,
our synthesis method works also as a proof method which is a decision
procedure for closed WS1S formulas.

If \( \varphi  \) is a \emph{closed} WS1S formula then the predicate
\( f \) introduced when constructing the set \( \mathit{Cls}(f,\varphi _{\tau }) \),
is a nullary predicate. Let \( \mathit{TransfP} \) be the program
derived by the unfold/fold transformation strategy starting from the
program \( \mathit{Nat}Set\cup \mathit{Cls}(f,\varphi _{\tau }) \).
As already known from Point (2) of Theorem~\ref{thm:corr_of_strategy},
we have that \( \mathit{Def}^{*}(f,\mathit{TransfP}) \) is either
the empty set or the singleton \( \{f\leftarrow \} \). Thus, we can
decide whether or not \( \mathcal{N}\models \varphi  \) holds by
checking whether or not \( f\leftarrow  \) belongs to \( \mathit{TransfP} \).
Since the unfold/fold transformation strategy always terminates, we
have that our unfold/fold synthesis method is indeed a decision procedure
for closed WS1S formulas. We summarize our proof method as follows.
\smallskip{}\newpage

\lyxline{\normalsize}\vspace{-1\parskip}
\noindent \textbf{The Unfold/Fold Proof} \textbf{\emph{}}\textbf{Method}\textbf{\emph{.}}

\noindent Let \( \varphi  \) be a closed WS1S formula. 

\noindent \emph{Step} 1. We apply the typed Lloyd-Topor transformation
and we derive the set \( \mathit{Cls}(f,\varphi _{\tau }) \) of clauses.

\noindent \emph{Step} 2. We apply the unfold/fold transformation strategy
and from the program \( \mathit{Nat}Set\cup \mathit{Cls}(f,\varphi _{\tau }) \)
we derive a definite program \( \mathit{TransfP} \).

\noindent If the unit clause \( f\leftarrow  \) belongs to \( \mathit{TransfP} \)
then \( \mathcal{N}\models \varphi  \) else \( \mathcal{N}\models \neg \varphi  \).
\lyxline{\normalsize}\smallskip{}

Now we present a simple example of application of our unfold/fold
proof method. 

\begin{example}
\label{ex:uf_proof}(\emph{An application of the unfold/fold proof
method.}) Let us consider the closed WS1S formula \( \varphi :\, \forall X\, \exists Y\, X\! \leq \! Y \).
By applying the typed Lloyd-Topor transformation starting from the
statement \( f\leftarrow \varphi  \), we get the following set of
clauses \( \mathit{Cls}(f,\varphi _{\tau }) \):
\smallskip{}

1. \( h(X)\leftarrow \mathit{nat}(X)\wedge \mathit{nat}(Y)\wedge X\! \leq \! Y \) 

2. \( g\leftarrow \mathit{nat}(X)\wedge \neg h(X) \) 

3. \( f\leftarrow \neg g \)
\smallskip{}

\noindent Now we apply the unfold/fold transformation strategy to
the program \( \mathit{Nat}Set \) and the following hierarchy of
natset-typed definitions: clause~1, clause~2, clause~3.

\noindent Initially, the program \( \mathit{TransfP} \) is \( \mathit{NatSet} \).
The transformation strategy proceeds left-to-right over that hierarchy.

\smallskip{}
\noindent (1) \( \mathit{Defs} \) and \( \mathit{InDefs} \) are
both set to \{clause 1\}. 

\noindent (1.1) \emph{Unfolding}. By unfolding, from clause 1 we get:
\smallskip{}

4. \( h(0)\leftarrow  \) 

5. \( h(0)\leftarrow \mathit{nat}(Y) \) 

6. \( h(s(X))\leftarrow \mathit{nat}(X)\wedge \mathit{nat}(Y)\wedge X\! \leq \! Y \) 
\smallskip{}

\noindent (1.2) \emph{Definition-Folding.} In order to fold the body
of clause 5 we introduce the following new clause:
\smallskip{}

7. \( \mathit{new}1\leftarrow \mathit{nat}(Y) \) 
\smallskip{}

\noindent Clause 6 can be folded by using clause 1. By folding clauses
5 and 6 we get:
\smallskip{}

8. \( h(0)\leftarrow \mathit{new}1 \) 

9. \( h(s(X))\leftarrow \mathit{h}(X) \) 
\smallskip{}

\noindent (1.3) At this point \( \mathit{TransfP}=\mathit{Nat}Set\cup \{ \)clause
4, clause 8, clause 9\}, \( \mathit{Defs}=\{ \)clause 1, clause 7\},
and \( \mathit{InDefs}=\{ \)clause 7\( \} \).

\noindent (1.4) By first unfolding clause 7 and then folding using
clause 7 itself, we get:
\smallskip{}

10. \( \mathit{new}1\leftarrow  \)

11. \( \mathit{new}1\leftarrow \mathit{new}1 \) 
\smallskip{}

\noindent No new clause is introduced (i.e., \( \mathit{NewDefs}=\emptyset  \)).
At this point \( \mathit{TransfP}=\mathit{Nat}Set\cup \{ \)clause
4, clause 8, clause 9, clause 10, clause 11\}, \( \mathit{Defs}=\{ \)clause
3, clause 7\}, and \( \mathit{InDefs}=\emptyset  \). Thus, the \noun{while}
statement terminates.

\noindent Since \( \mathit{Def}^{*}(\mathit{new}1,\mathit{TransfP}) \)
is propositional and \( M(\mathit{TransfP})\models \mathit{new}1 \),
by the propositional simplification rule we have: 

\( \mathit{TransfP}=\mathit{Nat}Set\cup \{ \)clause 4, clause 8,
clause 9, clause 10\( \} \).

\smallskip{}
\noindent (2) \( \mathit{Defs} \) is set to \{clause 1, clause 2,
clause 7\} and \( \mathit{InDefs} \) is set to \{clause 2\}. 

\noindent (2.1) \emph{Unfolding}. By unfolding, from clause 2 we get:
\smallskip{}

12. \( g\leftarrow \mathit{nat}(X)\wedge \neg h(X) \) 
\smallskip{}

\noindent (Notice that, by unfolding, clause \( g\leftarrow \neg h(0) \)
is deleted.)

\noindent (2.2) \emph{Definition-Folding.} Clause 12 can be folded
by using clause 2 which occurs in \( \mathit{Defs} \). Thus, no new
clause is introduced (i.e., \( \mathit{NewDefs}=\emptyset  \)) and
by folding we get:
\smallskip{}

13. \( g\leftarrow g \) 
\smallskip{}

\noindent (2.3) At this point \( \mathit{TransfP}=\mathit{Nat}Set\cup \{ \)clause
4, clause 8, clause 9, clause 10, clause 13\}, \( \mathit{Defs}=\{ \)clause
1, clause 2, clause 7\}, and \( \mathit{InDefs}=\emptyset  \). Thus,
the \noun{while} statement terminates. 

\noindent Since \( \mathit{Def}^{*}(g,\mathit{TransfP}) \) is propositional
and \( M(\mathit{TransfP})\models \neg g \), by the propositional
simplification rule we delete clause 13 from \( \mathit{TransfP} \)
and we have: 

\( \mathit{TransfP}=\mathit{Nat}Set\cup \{ \)clause 4, clause 8,
clause 9, clause 10\( \} \).

\smallskip{}
\noindent (3) \( \mathit{Defs} \) is set to \{clause 1, clause 2,
clause 3, clause 7\} and \( \mathit{InDefs} \) is set to \{clause
3\}. 

\noindent (3.1) \emph{Unfolding}. By unfolding clause 3 we get:
\smallskip{}

14. \( f\leftarrow  \) 
\smallskip{}

\noindent (Recall that, there is no clause in \( \mathit{TransfP} \)
with head \( g \).)

\noindent (3.2) \emph{Definition-Folding.} No transformation steps
are performed on clause~14 because it is a unit clause. 

\noindent (3.3) At this point \( \mathit{TransfP}=\mathit{Nat}Set\cup \{ \)clause
4, clause 8, clause 9, clause 10, clause 14\}, \( \mathit{Defs}=\{ \)clause
1, clause 2, clause 3, clause 7\}, and \( \mathit{InDefs}=\emptyset  \). 

\noindent The transformation strategy terminates and, since the final
program \( \mathit{TransfP} \) includes the unit clause \( f\leftarrow  \),
we have proved that \( \mathcal{N}\models \forall X\, \exists Y\, X\! \leq \! Y \). 

We would like to notice that neither SLDNF nor Tabled Resolution (as
implemented in the XSB system \cite{XSB00}) are able to construct
a refutation of \( \mathit{Nat}Set\cup \mathit{Cls}(f,\varphi _{\tau })\cup \{\leftarrow f\} \)
(and thus construct a proof of \( \varphi  \)), where \( \varphi  \)
is the WS1S formula \( \forall X\, \exists Y\, X\! \leq \! Y \).
Indeed, from the goal \( \leftarrow f \) we generate the goal \( \leftarrow \neg g \),
and neither SLDNF nor Tabled Resolution are able to infer that \( \leftarrow \neg g \)
succeeds by detecting that \( \leftarrow g \) generates an infinite
set of failed derivations. \hfill$\Box$
\end{example}
We would like to mention that some other transformations could be
applied for enhancing our unfold/fold transformation strategy. In
particular, during the strategy we may apply the subsumption rule
to shorten the transformation process by deleting some useless clauses.
For instance, in Example~\ref{ex:uf_proof} we can delete clause
5 which is subsumed by clause 4, thereby avoiding the introduction
of the new predicate \( \mathit{new}1 \). In some other cases we
can drop unnecessary type atoms. For instance, in Example~\ref{ex:uf_proof}
in clause~1 the type atom \( \mathit{nat}(X) \) can be dropped because
it is implied by the atom \( X\! \leq \! Y \). The program derived
at the end of the execution of the \noun{while} statement of the
unfold/fold transformation strategy are nondeterministic, in the sense
that an atom with non-variable arguments may be unifiable with the
head of several clauses. We can apply the technique for deriving deterministic
program presented in~\cite{Pe&97a} for deriving deterministic programs
and thus, obtaining smaller programs.

When the unfold/fold transformation strategy is used for program synthesis,
it is often the case that the above mentioned transformations also
improve the efficiency of the derived programs. 

Finally, we would like to notice that the unfold/fold transformation
strategy can be applied starting from a program \( P\cup \mathit{Cls}(f,\varphi _{\tau }) \)
(instead of \( \mathit{Nat}Set\cup \mathit{Cls}(f,\varphi _{\tau }) \))
where: (i) \( P \) is the output of a previous application of the
strategy, and (ii) \( \varphi  \) is a formula built like a WS1S
formula, except that it uses predicates occurring in \( P \) (besides
\( \leq  \) and \( \in  \)). Thus, we can synthesize programs (or
construct proofs) in a \emph{compositional} way, by first synthesizing
programs for subformulas. We will follow this compositional methodology
in the example of the following Section~\ref{sec:dbakery}.

\section{An Application to the Verification of Infinite State Systems: the
Dynamic Bakery Protocol \label{sec:dbakery}}

In this section we present an example of verification of a safety
property of an infinite state system by considering CLP(WS1S) programs~\cite{JaM94}.
As already mentioned, by applying our unfold/fold synthesis method
we will then translate CLP(WS1S) programs into logic programs.

The syntax of CLP(WS1S) programs is defined as follows. We consider
a set of \emph{user-defined} predicate symbols. A CLP(WS1S) clause
is of the form \( A\leftarrow \varphi \wedge G \), where \( A \)
is an atom, \( \varphi  \) is a formula of WS1S, \( G \) is a goal,
and the predicates occurring in \( A \) or in \( G \) are all user-defined.
A CLP(WS1S) \emph{program} is a set of CLP(WS1S) clauses. We assume
that CLP(WS1S) programs are \emph{}stratified. 

Given a CLP(WS1S) program \( P \), we define the semantics of \( P \)
to be its perfect model, denoted \emph{}\( M(P) \) (here we extend
to CLP(WS1S) programs the definitions which are given for normal logic
programs in~\cite{ApB94}). 

Our example concerns the Dynamic Bakery protocol, called \emph{DBakery}
for short, and we prove that it ensures mutual exclusion in a system
of processes which share a common resource, even if the number of
processes in the system changes during a protocol run in a dynamic
way. The \emph{DBakery} protocol is a variant of the \mbox{\emph{N-}process}
\emph{}Bakery protocol \cite{Lam74}.

In order to give the formal specifications of the \emph{DBakery} protocol
and its mutual exclusion property, we will use CLP(WS1S) as we now
indicate. The transition relation between pairs of system states,
the initial system state, and the system states which are \emph{unsafe}
(that is, the system states where more than one process uses the shared
resource) are specified by WS1S formulas. However, in order to specify
the mutual exclusion property we cannot use WS1S formulas only. Indeed,
mutual exclusion is a reachability property which is undecidable in
the case of infinite state systems. The approach we follow in this
example is to specify reachability (and, thus, mutual exclusion) as
a CLP(WS1S) program (see the program \( P_{\mathit{DBakery}} \) below).

Let us first describe the \emph{DBakery} protocol. We assume that
every process is associated with a natural number, called a \emph{counter},
and two distinct processes have distinct counters. At each instant
in time, the system of processes is represented by a pair \( \left\langle W,U\right\rangle  \),
called a \emph{system state}, where \( W \) is the set of the counters
of the processes \emph{waiting} for the resource, and \( U \) is
the set of the counters of the processes \emph{using} the resource.

A system state \( \left\langle W,U\right\rangle  \) is \emph{initial}
iff \( W\cup U \) is the empty set.

The transition relation from a system state \( \left\langle W,U\right\rangle  \)
to a new system state \( \left\langle W',U'\right\rangle  \) is the
union of the following three relations:

\smallskip{}
\noindent (T1: \emph{creation of a process}) 

\noindent \emph{~~~if} \( W\cup U \) is empty \emph{then} \( \left\langle W',U'\right\rangle =\left\langle \{0\},\emptyset \right\rangle  \)
\emph{else} \( \left\langle W',U'\right\rangle  \) \( = \) \( \left\langle W\cup \{m\! +\! 1\},\, U\right\rangle  \),

\noindent ~~~where \( m \) is the maximum counter in \( W\cup U \),

\smallskip{}
\noindent (T2: \emph{use of the resource}) 

\noindent \emph{~~~if} there exists a counter \( n \) in \( W \)
which is the minimum counter in \( W\cup U \) 

\noindent ~~~\emph{then} \( \left\langle W',U'\right\rangle =\left\langle W\! -\! \{n\},\, U\cup \{n\}\right\rangle  \),

\smallskip{}
\noindent (T3: \emph{release} \emph{of} \emph{the resource}) 

\noindent \emph{~~~if} there exists a counter \( n \) in \( U \)
\emph{then} \( \left\langle W',U'\right\rangle =\left\langle W,U\! -\! \{n\}\right\rangle  \).
\smallskip{}

\noindent The mutual exclusion property holds iff from the initial
system state it is not possible to reach a system state \( \left\langle W,U\right\rangle  \)
which is \emph{unsafe}, that is, such that \( U \) is a set of at
least two counters.
\medskip{}

Let us now give the formal specification of the \emph{DBakery} protocol
and its mutual exclusion property. We first introduce the following
WS1S formulas (between parentheses we indicate their meaning): 

\smallskip{}
\begin{tabular}{l}
\emph{empty}(\emph{X}) \( \, \equiv \,  \) \( \neg \exists x \)~\( x\! \in \! X \)\\
~~~~~~(the set \( X \) is empty)\vspace{1mm}\\
\emph{max}(\emph{X},\emph{m}) \( \, \equiv \,  \) \( m\! \in \! X \)~\( \wedge  \)~\( \forall x \)
(\( x\! \in \! X \)~\( \rightarrow  \)~\( x\! \leq \! m \)) \\
~~~~~~(\( m \) is the maximum in the set \( X \))\vspace{1mm}\\
\emph{min}(\emph{X},\emph{m}) \( \, \equiv \,  \) \( m\! \in \! X \)~\( \wedge  \)~\( \forall x \)
(\( x\! \in \! X \)~\( \rightarrow  \)~\( m\! \leq \! x \))\\
~~~~~~(\( m \) is the minimum in the set \( X \))\\
\end{tabular}
\smallskip{}

\noindent (Here and in what follows, for reasons of readability, we
allow ourselves to use lower case letters for individual variables
of WS1S formulas.) 

\noindent A system state \( \left\langle W,U\right\rangle  \) is
\emph{initial} iff \( \mathcal{N}\models \mathit{init}(\left\langle W,U\right\rangle ) \),
where:

\smallskip{}
\( \mathit{init}(\left\langle W,U\right\rangle ) \) \( \, \equiv \,  \)
\( \mathit{empty}(W) \) \( \wedge  \) \( \mathit{empty}(U) \)

\medskip{}
\noindent The transition relation \( R \) between system states is
defined as follows: 

\smallskip{}
\noindent \( \left\langle \left\langle W,U\right\rangle ,\, \left\langle W',U'\right\rangle \right\rangle \in R \)
iff 

\noindent \( \mathcal{N}\models \mathit{cre}(\left\langle W,U\right\rangle ,\left\langle W',U'\right\rangle ) \)
\( \vee  \) \( use(\left\langle W,U\right\rangle ,\left\langle W',U'\right\rangle ) \)
\( \vee  \) \( \mathit{rel}(\left\langle W,U\right\rangle ,\left\langle W',U'\right\rangle ) \)

\smallskip{}
\noindent where the predicates \emph{cre}, \emph{use}, and \emph{rel}
define the transition relations T1, T2, and T3, respectively. We have
that:

\smallskip{}
\begin{tabular}{l}
\( \mathit{cre}(\left\langle W,U\right\rangle ,\, \left\langle W',U'\right\rangle ) \)
\( \, \equiv \,  \) \( U'\! =\! U \)~\( \wedge  \)~\( \exists Z\, (Z\! =\! W\cup U\wedge  \)\\
\hspace{3.5cm}\( ((\mathit{empty}(Z)\wedge W'\! =\! \{0\})\, \vee  \)\\
\hspace{3.5cm}\( (\neg \mathit{empty}(Z)\wedge \exists m\, (\mathit{max}(Z,m)\wedge W'\! =\! W\! \cup \! \{s(m)\})))) \)\vspace{1.5mm}\\
\( use(\left\langle W,U\right\rangle ,\, \left\langle W',U'\right\rangle ) \)
\( \, \equiv \,  \) \( \exists n\, (n\in W\wedge \exists Z\, (Z\! =\! W\cup U\wedge \mathit{min}(Z,n))\, \wedge  \)\\
\hspace{3.5cm}\( W'\! =\! W\! -\! \{n\}\, \wedge \, U'\! =\! U\! \cup \! \{n\}) \)\vspace{1.5mm}\\
\( \mathit{rel}(\left\langle W,U\right\rangle ,\, \left\langle W',U'\right\rangle ) \)
\( \, \equiv \,  \) \( W'\! =\! W \)~\( \wedge  \)~\( \exists n\, (n\in U\wedge U'\! =\! U\! -\! \{n\}) \)\\
\end{tabular}
\smallskip{}

\noindent where the subformulas involving the set union \( (\cup ) \),
set difference \( (-) \), and set equality \( (=) \) operators can
be expressed as WS1S formulas.

Mutual exclusion holds in a system state \( \left\langle W,U\right\rangle  \)
iff \( \mathcal{N}\models \neg \mathit{unsafe}(\left\langle W,U\right\rangle ) \),
where \( \mathit{unsafe}(\left\langle W,U\right\rangle ) \) \( \, \equiv \,  \)
\( \exists n_{1}\, \exists n_{2}\, (n_{1}\! \in \! U\, \wedge \, n_{2}\! \in \! U\, \wedge \, \neg (n_{1}\! =\! n_{2})) \),
i.e., a system state \( \left\langle W,U\right\rangle  \) is unsafe
iff there exist at least two distinct counters in \( U \).

Now we will specify the system states reached from a given initial
system state by introducing the CLP(WS1S) program \( P_{\mathit{DBakery}} \)
consisting of the following clauses:
\smallskip{}

\( \mathit{reach}(S) \) \( \leftarrow  \) \( \mathit{init}(S) \)

\( \mathit{reach}(S1) \) \( \leftarrow  \) \( \mathit{cre}(S,S1) \)~\( \wedge  \)~\( \mathit{reach}(S) \)

\( \mathit{reach}(S1) \) \( \leftarrow  \) \( use(S,S1) \)~\( \wedge  \)~\( \mathit{reach}(S) \)

\( \mathit{reach}(S1) \) \( \leftarrow  \) \( \mathit{rel}(S,S1) \)~\( \wedge  \)~\( \mathit{reach}(S) \)

\smallskip{}
\noindent where \( \mathit{init}(S) \), \( \mathit{cre}(S,S1) \),
\emph{}\( use(S,S1) \), \emph{}and \emph{}\( \mathit{rel}(S,S1) \)
are the WS1S formulas listed above.

From \( P_{\mathit{DBakery}} \) we derive a definite program \( P'_{\mathit{DBakery}} \)
by replacing the WS1S formulas occurring in \( P_{\mathit{DBakery}} \)
by the corresponding atoms \( \mathit{init}(S) \), \( \mathit{cre}(S,S1) \),
\emph{}\( use(S,S1) \), \emph{}and \emph{}\( \mathit{rel}(S,S1) \),
and by adding to the program the clauses (not listed here) defining
these atoms, which are derived from the corresponding WS1S formulas
listed above, by applying the unfold/fold synthesis method (see Section~\ref{sec:synthesis}).
Let us call these clauses \emph{Init}, \emph{Cre}, \emph{Use}, and
\emph{Rel}, respectively. 
\medskip{}

In order to verify that the \emph{DBakery} protocol ensures mutual
exclusion for every system of processes whose number dynamically changes
over time, we have to prove that for every ground term \( \mathit{s} \)
denoting a finite set of counters, \( \mathit{ur}(\mathit{s})\not \in M(P'_{\mathit{DBakery}}\cup \{\rm {clause}\, 1\}) \),
where clause 1 is the following clause which we introduce by the definition
rule:

\smallskip{}
1.~\( \mathit{ur}(S) \) \( \leftarrow  \) \( \mathit{unsafe}(S) \)~\( \wedge  \)~\( \mathit{reach}(S) \) 
\smallskip{}

\noindent and \( \mathit{unsafe}(S) \) is defined by a set, called
\emph{Unsafe}, of clauses which are derived from the corresponding
WS1S formula by using the unfold/fold synthesis method.

In order to verify the mutual exclusion property for the \emph{DBakery}
protocol it is enough to show that \( P'_{\mathit{DBakery}}\cup \{\rm {clause}\, 1\} \)
can be transformed into a new definite program without clauses for
\( \mathit{ur}(S) \). This transformation can be done, as we now
illustrate, by a straightforward adaptation of the proof technique
presented for Constraint Logic Programs in \cite{Fi&02a}. In particular,
before performing folding steps, we will add suitable atoms in the
bodies of the clauses to be folded.

We start off this verification by unfolding clause~1 w.r.t.~the
atom \emph{reach}. We obtain the following clauses:

\smallskip{}
2.~\( \mathit{ur}(S) \) \( \leftarrow  \) \( \mathit{unsafe}(S) \)~\( \wedge  \)~\( \mathit{init}(S) \) 

3.~\( \mathit{ur}(S1) \) \( \leftarrow  \) \( \mathit{unsafe}(S1) \)~\( \wedge  \)~\( \mathit{cre}(S,S1) \)~\( \wedge  \)~\( \mathit{reach}(S) \)

4.~\( \mathit{ur}(S1) \) \( \leftarrow  \) \( \mathit{unsafe}(S1) \)~\( \wedge  \)~\( \mathit{use}(S,S1) \)~\( \wedge  \)~\( \mathit{reach}(S) \)

5.~\( \mathit{ur}(S1) \) \( \leftarrow  \) \( \mathit{unsafe}(S1) \)~\( \wedge  \)~\( \mathit{rel}(S,S1) \)~\( \wedge  \)~\( \mathit{reach}(S) \)

\smallskip{}
\noindent Now we can remove clause 2 because

\smallskip{}
\( M(\mathit{Unsafe}\cup \mathit{Init})\models \neg \exists S\, (\mathit{unsafe}(S)\wedge \mathit{init}(S)) \). 
\smallskip{}

\noindent The proof of this facts and the proofs of the other facts
we state below, are performed by applying the unfold/fold proof method
of Section~\ref{sec:synthesis}. Then, we fold clauses 3 and 5 by
using the definition clause~1 and we obtain:

\smallskip{}
6.~\( \mathit{ur}(S1) \) \( \leftarrow  \) \( \mathit{unsafe}(S1) \)~\( \wedge  \)~\( \mathit{cre}(S,S1) \)~\( \wedge  \)~\( \mathit{ur}(S) \)

7.~\( \mathit{ur}(S1) \) \( \leftarrow  \) \( \mathit{unsafe}(S1) \)~\( \wedge  \)~\( \mathit{rel}(S,S1) \)~\( \wedge  \)~\( \mathit{ur}(S) \)

\smallskip{}
\noindent Notice that this application of the folding rule is justified
by the following two facts:

\smallskip{}
\( M(\mathit{Unsafe}\cup \mathit{Cre}) \) \( \models \forall S\, \forall S1 \)~\( (\mathit{unsafe}(S1) \)~\( \wedge  \)~\( \mathit{cre}(S,S1) \)~\( \rightarrow  \)~\( \mathit{unsafe}(S)) \)

\( M(\mathit{Unsafe}\cup \mathit{Rel}) \) \( \models \forall S\, \forall S1 \)~\( (\mathit{unsafe}(S1) \)~\( \wedge  \)~\( \mathit{rel}(S,S1) \)~\( \rightarrow  \)~\( \mathit{unsafe}(S)) \)

\smallskip{}
\noindent so that, before folding, we can add the atom \( \mathit{unsafe}(S) \)
to the bodies of clauses 3 and 5. Now, since \( M(\mathit{Unsafe}\cup \mathit{Use})\models \neg \forall S\, \forall S1\, (\mathit{unsafe}(S1)\wedge \mathit{use}(S,S1) \)
\( \rightarrow  \) \( \mathit{unsafe}(S)) \), clause 4 \emph{cannot}
be folded using the definition clause~1. Thus, we introduce the new
definition clause:

\smallskip{}
8.\emph{~p}1\( (S) \) \( \leftarrow  \) \( c(S) \)~\( \wedge  \)~\( \mathit{reach}(S) \) 

\smallskip{}
\noindent where \( c(\left\langle W,U\right\rangle ) \) \( \equiv  \)
\( \exists n\, (n\! \in \! W\wedge \exists Z\, (Z=W\! \cup \! U\wedge \mathit{min}(Z,n))) \)
\( \wedge  \) \( \neg \mathit{empty}(U) \) which means that: in
the system state \( \left\langle W,U\right\rangle  \) there is at
least one process which uses the resource and there exists a process
waiting for the resource with counter \( n \) which is the minimum
counter in \( W\cup U \). 

Notice that, by applying the unfold/fold synthesis method, we may
derive a set, called \emph{Busy} (not listed here), of definite clauses
which define \( c(S) \).

\noindent By using clause 8 we fold clause 4, and we obtain: 

\smallskip{}
9.~\( \mathit{ur}(S1) \) \( \leftarrow  \) \( \mathit{unsafe}(S1) \)~\( \wedge  \)~\( use(S,S1) \)~\( \wedge  \)~\emph{p}1\( (S) \)
\smallskip{}

\noindent We proceed by applying the unfolding rule to the newly introduced
clause 8, thereby obtaining:

\smallskip{}
10.~\emph{p}1\( (S) \) \( \leftarrow  \) \( \mathit{c}(S) \)~\( \wedge  \)\emph{~}\( \mathit{init}(S) \) 

11.\emph{~p}1\( (S1) \) \( \leftarrow  \) \( c(S1) \)~\( \wedge  \)~\( \mathit{cre}(S,S1) \)~\( \wedge  \)~\( \mathit{reach}(S) \) 

12.\emph{~p}1\( (S1) \) \( \leftarrow  \) \( c(S1) \)~\( \wedge  \)~\( use(S,S1) \)~\( \wedge  \)~\( \mathit{reach}(S) \) 

13.\emph{~p}1\( (S1) \) \( \leftarrow  \) \( c(S1) \)~\( \wedge  \)~\( \mathit{rel}(S,S1) \)~\( \wedge  \)~\( \mathit{reach}(S) \) 
\smallskip{}

\noindent Clauses 10 and 12 are removed, because

\smallskip{}
\( M(\mathit{Busy}\cup \mathit{Init})\models \neg \exists S\, (c(S)\wedge \mathit{init}(S)) \)

\( M(\mathit{Busy}\cup \mathit{Use})\models \neg \exists S\, \exists S1\, (c(S1)\wedge \mathit{use}(S,S1)) \)
\smallskip{}

\noindent We fold clauses 11 and 13 by using the definition clauses
8 and 1, respectively, thereby obtaining:

\smallskip{}
14.\emph{~p}1\( (S1) \) \( \leftarrow  \) \( c(S1) \)~\( \wedge  \)~
\( \mathit{cre}(S,S1) \)~\( \wedge  \)~\( p1(S) \) 

15.\emph{~p}1\( (S1) \) \( \leftarrow  \) \( c(S1) \)~\( \wedge  \)~
\( \mathit{rel}(S,S1) \)~\( \wedge  \)~\( \mathit{ur}(S) \) 

\smallskip{}
\noindent Notice that this application of the folding rule is justified
by the following two facts:

\smallskip{}
\( M(\mathit{Busy}\cup \mathit{Cre})\models  \) \( \forall S\, \forall S1 \)~\( ((c(S1) \)~\( \wedge  \)~\( \mathit{cre}(S,S1)) \)~\( \rightarrow  \)~\( c(S)) \)

\( M(\mathit{Busy}\cup \mathit{Rel})\models  \) \( \forall S\, \forall S1 \)~\( ((c(S1) \)~\( \wedge  \)~\( \mathit{rel}(S,S1)) \)~\( \rightarrow  \)~\( \mathit{unsafe}(S)) \)

\smallskip{}
\noindent Thus, starting from program \( P'_{\mathit{DBakery}}\cup  \)\{clause~1\}
we have derived a new program \( Q \) consisting of clauses 6, 7,
14, and 15. Since all clauses in \( \mathit{Def}^{*}(\mathit{ur},Q) \)
are recursive, we have that for every ground term \( \mathit{s} \)
denoting a finite set of counters, \( \mathit{ur}(\mathit{s})\not \in M(Q) \)
and by the correctness of the transformation rules~\cite{PeP02a},
we conclude that mutual exclusion holds for the \emph{DBakery} protocol.

\section{Related Work and Conclusions}

We have proposed an automatic synthesis method based on unfold/fold
program transformations for translating CLP(WS1S) programs into normal
logic programs. This method can be used for avoiding the use of ad-hoc
solvers for WS1S constraints when constructing proofs of properties
of infinite state multiprocess systems. 

Our synthesis method follows the general approach presented in \cite{PeP02a}
and it terminates for any given WS1S formula. No such termination
result was given in \cite{PeP02a}. In this paper we have also shown
that, when we start from a closed WS1S formula \( \varphi  \), our
synthesis strategy produces a program which is either (i) a unit clause
of the form \( f\leftarrow  \), where \( f \) is a nullary predicate
equivalent to the formula \( \varphi  \), or (ii) the empty program.
Since in case (i) \( \varphi  \) is true and in case (ii) \( \varphi  \)
is false, our strategy is also a decision procedure for closed WS1S
formulas. This result extends \cite{PeP00a} which presents a decision
procedure based on the unfold/fold proof method for the \emph{clausal
fragment} of the WSkS theory, i.e., the fragment dealing with universally
quantified disjunctions of conjunctions of literals. 

Some related methods based on program transformation have been recently
proposed for the verification of infinite state systems \cite{LeM99,RoR01}.
However, as it is shown by the example of Section~\ref{sec:dbakery},
an important feature of our verification method is that the number
of processes involved in the protocol may change over time and other
methods find it problematic to deal with such dynamic changes. In
particular, the techniques presented in \cite{RoR01} for verifying
safety properties of \emph{parametrized systems} deal with reactive
systems where the number of processes is a parameter which does not
change over time.

Our method is also related to a number of other methods which use
logic programming and, more generally, constraint logic programming
for the verification of reactive systems (see, for instance,~\cite{DeP99,FrO97b,NiL00,Ra&97}
and~\cite{Fri00} for a survey). The main novelty of our approach
w.r.t.~these methods is that it combines logic programming and monadic
second order logic, thereby modelling in a very direct way systems
with an unbounded (and possibly variable) number of processes.

Our unfold/fold synthesis method and our unfold/fold proof method
have been implemented by using the MAP transformation system~\cite{MAP}.
Our implementation is reasonably efficient for WS1S formulas of small
size (see the example formulas of Section~\ref{sec:dbakery}). However,
our main concern in the implementation was not efficiency and our
system should not be compared with ad-hoc, well-established theorem
provers for WS1S formulas based on automata theory, like the MONA
system \cite{He&96}. Nevertheless, we believe that our technique
has its novelty and deserves to be developed because, being based
on unfold/fold rules, it can easily be combined with other techniques
for program derivation, specialization, synthesis, and verification,
which are also based on unfold/fold transformations. 

\bibliographystyle{plain}
\bibliography{Transformation}

\end{document}